\documentclass[
    pre,
    twocolumn,
    twoside,
    superscriptaddress,
    floatfix,
    nofootinbib,
    longbibliography
]{revtex4-1}

\usepackage[realmainfile]{currfile}
\usepackage{url}
\PassOptionsToPackage{hyphens}{url}\usepackage{hyperref}
\makeatletter
\g@addto@macro{\UrlBreaks}{\UrlOrds}
\makeatother

\usepackage{hyperref}
\hypersetup{
  colorlinks=true,
  allcolors=todoblue,
  urlcolor=todoblue,
  citecolor=todoblue,
  pdfborder={0 0 0},
  breaklinks=true,
}

\usepackage{colortbl}

\makeatletter

\def\CT@@do@color{%
  \global\let\CT@do@color\relax
  \@tempdima\wd\z@
  \advance\@tempdima\@tempdimb
  \advance\@tempdima\@tempdimc
  \advance\@tempdimb\tabcolsep
  \advance\@tempdimc\tabcolsep
  \advance\@tempdima2\tabcolsep
  \kern-\@tempdimb
  \leaders\vrule
  \hskip\@tempdima\@plus  1fill
  \kern-\@tempdimc
  \hskip-\wd\z@ \@plus -1fill }
\makeatother

\usepackage{xcolor}

\definecolor{olivegreen}{rgb}{0.33333,.41961,0.18431}
\definecolor{forestgreen}{rgb}{0.13333,.5451,0.13333}
\definecolor{lightgrey}{rgb}{0.7,0.7,0.7}
\definecolor{verylightgrey}{rgb}{0.90,0.90,0.90}
\definecolor{veryverylightgrey}{rgb}{0.95,0.95,0.95}
\definecolor{grey}{rgb}{0.5,0.5,0.5}

\definecolor{headerblue}{HTML}{33367E}
\definecolor{unitednationsblue}{HTML}{4D88FF}

\definecolor{charcoal}{HTML}{36454F}
\definecolor{cinerous}{HTML}{98817B}
\definecolor{feldgrau}{HTML}{4D5D53}
\definecolor{glaucous}{HTML}{6082B6}
\definecolor{arsenic}{HTML}{3B444B}
\definecolor{xanadu}{HTML}{738678}

\definecolor{firebrick}{HTML}{B22222}
\definecolor{orangered}{HTML}{FF4500}
\definecolor{tomato}{HTML}{FF6347}

\definecolor{purpletaupe}{HTML}{3B444B}
\definecolor{todoblue}{RGB}{0, 91, 187}

\usepackage{graphicx,epsfig,verbatim,enumerate}
\usepackage{amssymb,amsmath}
\usepackage{ifthen}

\usepackage{longtable}

\usepackage{tabularx}
\usepackage{multirow, makecell}
\newcolumntype{C}{>{\centering\arraybackslash}X}
\newcolumntype{L}{>{\raggedright\arraybackslash}X}
\newcolumntype{R}{>{\raggedleft\arraybackslash}X}
\usepackage{mathtools}

\newboolean{twocolswitch}

\newcommand{\sindex}[1]{}
\newcommand{\nindex}[1]{}

\newcommand{\www}[1]{\url{#1}}

\usepackage{lettrine}

\newcommand{\ngramprop}{p_{\tau,t}}

\setboolean{twocolswitch}{true}

\renewcommand{\thesection}{\arabic{section}}
\renewcommand{\thesubsection}{\arabic{section}.\arabic{subsection}}

\renewcommand{\thefigure}{\textbf{\arabic{figure}}}

\renewcommand{\thetable}{\textbf{\arabic{table}}}

\makeatletter
\def\p@subsection{}
\makeatother

\begin{document}

\title{\protect
  Say Their Names: 
Resurgence in the collective attention 
toward Black victims of 
fatal police violence 
following the death of George Floyd
}

\author{
\firstname{Henry H.}
\surname{Wu}
}
%\email{henry.wu@uvm.edu}

\affiliation{
    Computational Story Lab,
    MassMutual Center of Excellence for Complex Systems and Data Science,
    Vermont Advanced Computing Core, University of Vermont, Burlington, VT 05401.
}

\affiliation{
    Vermont Complex Systems Center, University of Vermont, Burlington, VT 05401.
}

\author{
\firstname{Ryan J.}
\surname{Gallagher}
}

\affiliation{
  Communication Media and Marginalization Lab, Network Science Institute, Northeastern University, Boston, MA, 02115
}

\author{
\firstname{Thayer}
\surname{Alshaabi}
}

\affiliation{
    Computational Story Lab,
    MassMutual Center of Excellence for Complex Systems and Data Science,
    Vermont Advanced Computing Core, University of Vermont, Burlington, VT 05401.
}

\affiliation{
    Vermont Complex Systems Center, University of Vermont, Burlington, VT 05401.
}

\author{
\firstname{Jane L.}
\surname{Adams}
}

\affiliation{
    Computational Story Lab, 
    MassMutual Center of Excellence for Complex Systems and Data Science,
    Vermont Advanced Computing Core, University of Vermont, Burlington, VT 05401.
}

\affiliation{
    Vermont Complex Systems Center, University of Vermont, Burlington, VT 05401.
}

\author{
\firstname{Joshua R.}
\surname{Minot}
}

\affiliation{
    Computational Story Lab, 
    MassMutual Center of Excellence for Complex Systems and Data Science,
    Vermont Advanced Computing Core, University of Vermont, Burlington, VT 05401.
}

\affiliation{
    Vermont Complex Systems Center, University of Vermont, Burlington, VT 05401.
}
\author{
\firstname{Michael V.}
\surname{Arnold}
}

\affiliation{
    Computational Story Lab, 
    MassMutual Center of Excellence for Complex Systems and Data Science,
    Vermont Advanced Computing Core, University of Vermont, Burlington, VT 05401.
}

\affiliation{
    Vermont Complex Systems Center, University of Vermont, Burlington, VT 05401.
}

\author{
\firstname{Brooke}
\surname{Foucault Welles}
}

\affiliation{
  Communication Media and Marginalization Lab, Network Science Institute, Northeastern University, Boston, MA, 02115
}
\affiliation{
  Communication Studies Department, Northeastern University, Boston, MA, 02115
}

\author{
\firstname{Randall}
\surname{Harp}
}

\affiliation{
    Vermont Complex Systems Center, University of Vermont, Burlington, VT 05401.
}

\affiliation{
    Department of Philosophy, 
    University of Vermont, Burlington, VT 05401.
}

\author{
\firstname{Peter}
\surname{Sheridan Dodds}
}

\email{peter.dodds@uvm.edu}

\affiliation{
    Computational Story Lab, 
    MassMutual Center of Excellence for Complex Systems and Data Science,
    Vermont Advanced Computing Core, 
    University of Vermont, Burlington, VT 05401.
}

\affiliation{
    Vermont Complex Systems Center, University of Vermont, Burlington, VT 05401.
}

\affiliation{Department of Computer Science, University of Vermont, Burlington, VT 05401.}

\author{
\firstname{Christopher}
\surname{M. Danforth}
}

\email{chris.danforth@uvm.edu}

\affiliation{
    Computational Story Lab, 
    MassMutual Center of Excellence for Complex Systems and Data Science,
    Vermont Advanced Computing Core, University of Vermont, Burlington, VT 05401.
}

\affiliation{
    Vermont Complex Systems Center, University of Vermont, Burlington, VT 05401.
}

\affiliation{Department of Mathematics \& Statistics, University of Vermont, Burlington, VT 05401.}

\date{\today}

\begin{abstract}
  \protect
  The murder of George Floyd by police in May 2020 sparked international protests and renewed attention in the Black Lives Matter movement. 
Here, we characterize ways in which the online activity following George Floyd's death was unparalleled in its volume and intensity, including setting records for activity on Twitter, 
prompting the saddest day in the platform’s history,
and causing George Floyd’s name to appear among the ten most frequently used phrases in a day, 
where he is the only individual to have ever received that level of attention who was not known to the public earlier that same week. 
Further, we find this attention extended beyond George Floyd and that more Black victims of fatal police violence received attention following his death than during other past moments in Black Lives Matter’s history. 
We place that attention within the context of prior online racial justice activism by showing how the names of Black victims of police violence have been lifted and memorialized over the last 12 years on Twitter. 
Our results suggest that the 2020 wave of attention to the Black Lives Matter movement centered past instances of police violence in an unprecedented way, demonstrating the impact of the movement’s rhetorical strategy to ``say their names.’’ 
\end{abstract}


\maketitle

% =============================================================
% =============================================================
% ======================= Introduction ========================
% =============================================================
% =============================================================
\section{Introduction}
\label{sec:introduction}

    On February 23rd, 2020, Ahmaud Arbery, a 25-year-old Black man, was shot and killed by three white men while jogging in Georgia.
    On March 13th, Breonna Taylor, a 26-year-old Black woman, was fatally shot in the crossfire of a ``no-knock’’ apartment search by police in Kentucky. 
    And on May 25th, George Floyd, a 46-year-old Black man, was arrested outside a convenience store in Minnesota, and murdered when white police officer Derek Chauvin knelt on his neck for over 9 minutes. 
    Catalyzing the anger and grief that had been circulating the deaths of Arbery and Taylor, Floyd's murder sparked protests across the United States, bringing renewed attention to police brutality and racism. 
    The protests were coupled with an unprecedented use of the hashtag \#BlackLivesMatter \cite{anderson2020black}, surpassing every previous surge of the hashtag since its introduction in July 2013 following the acquittal of George Zimmerman in the death of Trayvon Martin and its widespread adoption in November 2014 following the non-indictment of police officer Darren Wilson in the death of Michael Brown \cite{garza2014herstory,freelon2016beyond}. It is estimated that 15 to 26 million Americans participated in racial justice protests in June 2020, making them the largest protests in American history~\cite{buchanan2020black}. 

    The Black Lives Matter movement has connected many instances of police violence against Black individuals, including Ahmaud Arbery, Breonnna Taylor, and George Floyd, into a cohesive narrative about systemic racism in the United States. In December 2014, the African American Policy Forum and the Center for Intersectionality and Social Policy Studies launched the \#SayHerName campaign, which ``brings awareness to the often invisible names and stories of Black women and girls who have been victimized by racist police violence, and provides support to their families''.\footnote{\url{https://www.aapf.org/sayhername}} Since then, the related phrase ``say their names’’ and the hashtag \#SayTheirNames have been invoked to recognize Black victims of police violence more broadly. While some have critiqued \#SayTheirNames for drawing attention away to the focus on women and girls (\#SayHerName), repeating the names of  victims of police violence of all genders serves an important narrative function. Naming victims memorializes and celebrates these individuals, while emphasizing their place in a larger system of police violence and racial prejudice~\cite{jackson2020hashtagactivism}. This rhetorical strategy contributes to the goals of Black Lives Matter activists, which include ``education, amplification of marginalized voices, and structural police reform''~\cite{freelon2016beyond}.

    Here, we characterize the ways in which the level of online attention given to George Floyd’s death and the subsequent protests on Twitter was unprecedented. In the wake of Floyd’s death, there was an unparalleled surge in tweet volume driven by historic levels of retweet amplification, which coincided with a substantial and sustained dip in the happiness expressed on Twitter. Of the few dozen individuals who have ever received comparable amounts of attention online, George Floyd is the only person who was not already a celebrity. Critically though, attention was not given \emph{only} to George Floyd, nor to the protests sparked by his murder. By observing how 3,737 names were used during and after the protests, we show that there was also a resurgence in attention to past instances of fatal police violence against Black Americans. This resurgence was instantaneous and more persistent than previous spikes in attention to the Black Lives Matter movement. Recognizing that the resurgence of attention built upon years of movement building and racial justice activism, we examine how those names have been given attention and amplified over the past 12 years on Twitter. Across the past decade, we see patterns of increasing attention to and amplification of Black victims of fatal police violence, and we see discernible trends in how often and widely different names have been used. Our study goes beyond the most emblematic names of the Black Lives Matter movement and shows the extent to which it is built upon thousands of often less visible, but no less important instances of people saying the names of Black victims of police violence.

% =============================================================
% =============================================================
% ======================= Related Work ========================
% =============================================================
% =============================================================

\section{Related Work}
\label{sec:relatedwork}
    
    The Black Lives Matter movement and its corresponding hashtag \#BlackLivesMatter are effective in part because they bring together otherwise isolated instances of police violence into a unified narrative. Bringing those experiences together is made easier by social media because it allows ordinary individuals---rather than political elites, journalists, and celebrities---to connect with one another and reach large audiences \cite{ellison2013sociality,benkler2008wealth}. Prior to social media, those audiences were difficult for social justice campaigns to reach directly; one of the only effective routes for doing so was through print and television news media, which often bar protest messaging \cite{vos2019journalists,entman1994representation}. Instead, social media gives activists, on-the-ground protesters, citizen journalists, and others a way to bypass traditional top-down structures of information dissemination and directly share their experiences with the public \cite{jackson2015hijacking}. Exactly because social media---and Twitter in particular---are valuable sources of information during rapidly evolving events, such as protests, journalists often look to online conversations to understand emerging events \cite{molyneux2021legitimating}. Rather than having to appeal directly to print and news media then, activists and others can find their messaging distributed more broadly in the public sphere when it is picked up by journalists via social media \cite{chadwick2011political,mcgregor2019social}. From the Indignados movement \cite{anduiza2014mobilization,theocharis2015using}, Occupy movement \cite{tremayne2014anatomy,conover2013digital}, and Arab Spring \cite{steinert2015online,tufekci2017twitter} of the early 2010s to the Umbrella movement \cite{lee2015social,lee2016digital} and \#MeToo campaign \cite{gallagher2019reclaiming,suk2021metoo} of the latter half of the decade, there is now a well-established line of protests whose prominence rested, in part, on their ability to redirect the attention of journalists and the public more widely through social media.

    The success of online activism is not simply spontaneous though. In contrast to offline collective action where organizing is the primary means through which protests emerge, online \emph{connective} action is often strung together through narratives that resonate with experiences of oppression and marginalization \cite{bennett2012logic,papacharissi2016affective}. Most social media provide a particularly effective mechanism for stringing together those stories: hashtags. On a purely technical level, hashtags connect different posts together through a common tag. In doing so, they provide a common banner under which people can share their stories, including those that are not often given attention in the public sphere \cite{jackson2015hijacking}. This allows \emph{counterpublic} spheres to emerge, online assemblages of people and posts that both reaffirm the lived experiences of marginalization and challenge mainstream narratives \cite{jackson2020hashtagactivism,squires2002rethinking}. By weaving individual stories together, counterpublic spaces organized around hashtags transform personal experiences into a larger network of connective action \cite{bennett2012logic,gallagher2019reclaiming}. In turn, this causes a protest hashtag to become more than just a tag in a post: it becomes a \emph{signifier} of a broader idea, feeling, or movement \cite{papacharissi2016affective}. The hashtag \#BlackLivesMatter, for example, does not just stand for the death of Trayvon Martin, for whom it was originally invented, nor Michael Brown, for whom the hashtag gained widespread visibility. It signifies a recognition of all past and future Black victims of police violence; it signifies a commitment to addressing violence against Black communities more broadly; and it signifies attention to persistent and historical systemic racism, particularly in the United States.

    Hashtags are not the only signifiers around which networked counterpublics can emerge, though. This is apparent in the phrases ``say her name’’ and ``say their names’’ in the Black Lives Matter movement. They both explicitly encourage people to say, repeat, and amplify the names of Black victims of fatal police violence. This is an important and notable portion of Black Lives Matter’s discursive strategy \cite{freelon2016beyond,ince2017social,jackson2020hashtagactivism,gallagher2018divergent}. Even though each new instance of extrajudicial police violence is very often not \emph{directly} related to a past incident, the names of past victims are still often reiterated online in the wake of a new victim’s death. This connects that instance to previous instances of police violence. By repeatedly saying the names of past victims of police violence, the names themselves become signifiers of the same topics, issues, and narratives that are signified by \#BlackLivesMatter more generally \cite{jackson2020hashtagactivism}. 

    Given the historic levels of attention given to the Black Lives Matter movement following the death of George Floyd \cite{anderson2020black,buchanan2020black}, there is a sense that the summer of 2020 was a pivotal moment for racial justice.  That moment was not focused only on Floyd, though: the deaths of Ahmaud Arbery and Breonna Taylor earlier that year played a notable role in the ensuing conversations. We focus on answering three research questions related to the intense attention that was given to George Floyd’s death, and how that moment invoked the names of past victims to position itself with respect to a longer history of police violence:
    \begin{enumerate}
        \item In what ways was George Floyd’s death and the following protests an exceptional moment in how racial justice was discussed on Twitter?
        \item How did people use the ``say their names’’ strategy to connect George Floyd to past Black victims of police violence?
        \item In what ways does the attention given to George Floyd build off a longer history of how the names of Black victims have been given attention on Twitter?
    \end{enumerate}

% ========================================================
% ========================================================
% ======================= Methods ========================
% ========================================================
% ========================================================
\section{Data and Methods}
\label{sec:data}
    
% --------------------------------------------------------------------
% -------------------- Methods: Fatal Encounters ---------------------
% --------------------------------------------------------------------
    \subsection{Police-Involved Deaths of Black Victims}
    \label{sec:policeviolence}
    
    Significant attention has been given to the deaths of Michael Brown, Sandra Bland, Philando Castile, Breonna Taylor, George Floyd, and other Black victims of police violence. Unfortunately, their cases are the exception, rather than the norm: most victims of police violence never receive considerable attention on social media, if any. So although it is important to characterize those incidents that have become emblematic of the Black Lives Matter movement, centering our analysis around \emph{only} them would overlook the many cases of police violence against Black communities that never made their way into mainstream conversations. 

    Instead, we start from a list of police-involved deaths more broadly and use that to measure the attention that has been given to the names of Black victims. Because of the irregular reporting of police-involved deaths~\cite{strom2017future, desmond2016police}, there is no official, complete accounting of victims of police violence, and federal crime data lacks the granularity that is needed to analyze specific incidents of police violence. To address this, we draw from the Fatal Encounters database, a third-party database of people killed during interactions with police officers. It contains records of over 29,000 people killed by police officers since 2000, documented through a mix of web scraping, manual investigation, public records requests, and crowdsourcing by paid researchers and volunteers~\cite{finch2019using}. It includes both Black and non-Black victims of police violence, and includes the date of the incident that resulted in their death and the cause of death.
    
    To align the Fatal Encounters database with our social media data, we consider deaths occurring from January 1st, 2009 onward. We focus specifically on the deaths of Black victims. To direct our analysis towards deaths that are directly caused by police actions, we exclude suicides and vehicular deaths. We also establish several criteria for inclusion based on the names of the victims. First, we remove names that received a measurable amount of attention in the 10 days \emph{prior} to the death of a Black victim with that same name (e.g. Michael Myers, George Bush). 
    We say that a name received \emph{measurable attention} when it was among the top million 2-grams for a day, and provide more detail in the next section. Next, we also manually remove 12 additional names, primarily those shared with famous athletes and two names shared with police officers involved in high-profile deaths related to Black Lives Matter (Darren Wilson and Thomas Lane). For duplicate names in the Fatal Encounters database, we attribute all mentions of a name to the earliest incident that pertains to that name. In Tables~\ref{tab:duplicates}-~\ref{tab:unknowns}, we list all of the names that were excluded from the analysis due to these steps. Finally, we also manually add 15 names significant to the Black Lives Matter movement that were not in the Fatal Encounters database, such as those that did not involve police (e.g. Trayvon Martin, Ahmaud Arbery) or those that occurred before database’s timeframe (e.g. Emmett Till, Rodney King). Conducting our analyses without these names would result in an incomplete picture of online attention to anti-Black violence in the United States. We provide the full list of manually included names in Table~\ref{tab:manual}. Together, these preprocessing steps yield 3,737 records.

% -----------------------------------------------------------------
% -------------------- Methods: Storywrangler ---------------------
% -----------------------------------------------------------------
    \subsection{Mentions of Victims' Names on Twitter}
    \label{sec:twitter}
    
    Because names are important signifiers in the Black Lives Matter movement \cite{freelon2016beyond,ince2017social,jackson2020hashtagactivism,gallagher2018divergent}, we measure how much attention has been given to the names of the victims we identified from the Fatal Encounters database since their deaths. We use the first and last name of each victim in our dataset to create a set of two word phrases, or 2-grams. For each name, we query its frequency and rank over time using the Storywrangler API\footnote{\url{https://storywrangling.org/}} \cite{alshaabi2020storywrangler}. Storywrangler uses a 10\% sample of English tweets to measure how often words and phrases (also known generally as $n$-grams\footnote{Note, $n$-grams are more general than $n$-word phrases split by whitespace. For example, 2-grams also include consecutive punctuation (e.g. ``!!'', ``??'') and combinations of punctuation and words (e.g. ``. and'', ``? I''). See ref.~\cite{alshaabi2020storywrangler} for the Storywrangler API's full definition of an $n$-gram.}) are used on Twitter. For a given $n$-gram, Storywrangler returns a time series of its frequency over time and a time series of its frequency rank over time. The rank is computed for each day by comparing how often a word was used compared to all other $n$-grams of the same length for that day. We retrieve these time series for each victim's name, starting from the date of their injury that resulted in death as recorded in the Fatal Encounters database.

    Importantly, Storywrangler only indexes time series data when a particular $n$-gram was among the top million most frequently used $n$-grams of that day. Of the 3,737 names we transcribed from the Fatal Encounters database, 2,603 (69.6\%) were never used often enough---neither on the day of their death nor on any day following---to rank as the one millionth most frequently used 2-gram on Twitter or higher. For deaths occurring in 2010, this threshold was roughly 20 mentions on a single day, and in 2020 the threshold was closer to 200 mentions.

       \begin{table}[!tp]
        \centering
        \begin{tabular}{c|c|c|c}
            Tweets (millions) & May 1--26 & May 27--June 7 & \% Change\\
            \hline
            \hline
            Original & 69.7 & 71.2 & +2.1\\
            \hline
            Retweets & 87.5 & 126 & +44\\
            \hline
            Total & 157 & 197 & +25\\
        \end{tabular}
        \caption{
            \textbf{Change in average daily tweet volume in May 2020 and the subsequent spike period by tweet type.} Messages are separated into originally authored tweets, retweets, and their sum (in millions) between May 1--26 and May 27--June 7, 2020. 
        }
        \label{tab:spike}
    \end{table}

   Unless stated otherwise, in the following analyses we always characterize the attention given to those who ranked within the top million 2-grams. We say those individuals received \emph{measurable attention.} While this subset is significantly larger than the emblematic names of Black Lives Matter---a fact which allows us to analyze how names are used in the movement in greater detail than we would be able to otherwise---we emphasize that the definitive majority of Black victims of police violence are never mentioned widely on Twitter.

    In addition to the time series data for the names, we collect similar data for the phrase ``Black Lives Matter'' and the hashtag \#BlackLivesMatter. We couple that data with measurements made by the Hedonometer, a dictionary-based instrument designed to provide a macro-level approximation of the ``happiness'' expressed in tweets~\cite{dodds2011temporal}. To better interpret how the expressed happiness changed in the wake of George Floyd's death, we also use word shift graphs \cite{gallagher2021generalized} to measure how particular words contributed to fluctuations in the expressed happiness compared to the previous week. See the Supplementary Materials for further detail on the sentiment analysis methods.
    
        \begin{figure*}[!tp]
          \centering	
            \includegraphics[width=\textwidth]{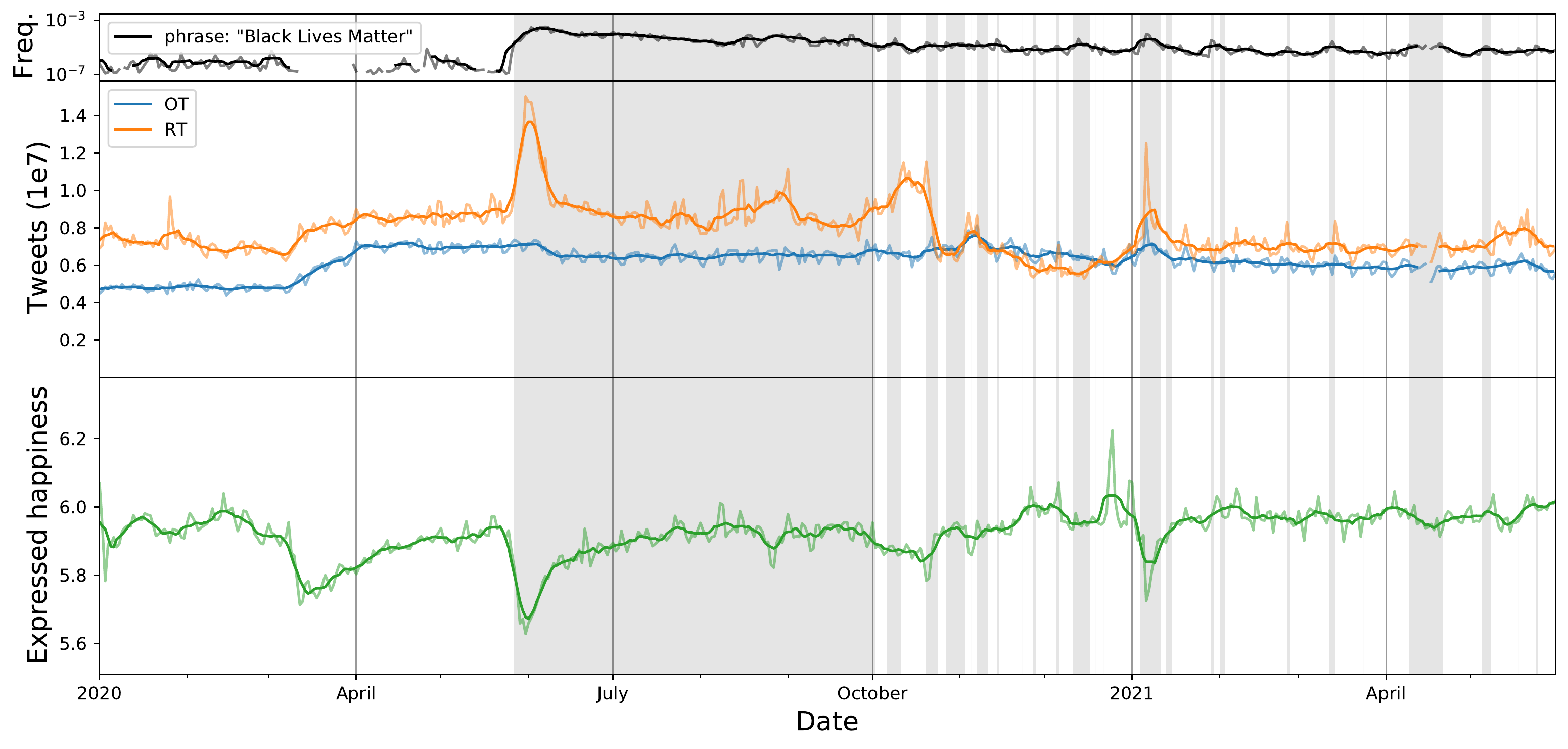}  
          \caption{
            \textbf{English tweet volume and expressed happiness (random 10\% sample), 2020-2021.} \textbf{Top)} Relative frequency of the phrase ``Black Lives Matter''. The time series is absent at times where ``Black Lives Matter'' did not receive measurable attention, i.e. it was not within the top million 3-grams. The gray shaded areas represent periods of time when ``Black Lives Matter" was among the top 5,000 most used 3-grams per day. 
            \textbf{Middle)} Daily counts for originally authored tweets (blue, OT) and retweets (orange, RT) reveal a spike period in late May and early June 2020 during which retweet activity set all time records.
            The sudden decrease in retweets in October 2020 is attributable to a platform-wide design change to retweets on Twitter \cite{gadde2020additional}.
            \textbf{Bottom)} Average expressed happiness of English-language Twitter shown by day. Drops in expressed happiness corresponding to the pandemic (March 2020) and the Capitol insurrection (January 2021) are both apparent.
            Less happiness was expressed in the period following George Floyd's death than any other point in Twitter's history. For all panels, light lines indicate the raw frequency and sentiment values, and darker bold lines indicate 7-day rolling averages.
          }
          \label{fig:spike}
    \end{figure*}

% --------------------------------------------------------------------------
% -------------------- Methods: Measures of Attention ----------------------
% --------------------------------------------------------------------------
    \subsection{Measures of Attention and Amplification}
    
    We use the frequency and rank time series from Storywrangler to define several different measures of collective attention and amplification. We start by noting that raw frequencies of names are subject to fluctuations in how many tweets were written in general on Twitter, which can vary widely in both short- and long-term time windows. Instead, for each day we calculate each name's relative frequency, which is the name's frequency normalized by the total frequency of all words on Twitter for a particular day---where we note that for all of the measures described below, we always compare to $n$-grams of the same length. So, that is, if $f_{\tau,t}$ is the raw frequency of a name $\tau$ on a given day $t$, then its relative frequency $p_{\tau,t}$ is
    \begin{equation}
        p_{\tau,t}
            =
                \frac{
                    f_{\tau,t}
                }{
                    \sum_{\tau' \in \mathcal{W}_t} f_{\tau', t}
                },
    \end{equation}
    where $\mathcal{W}_t$ is the collection of all $2$-grams used on that day. If a name does not receive measurable attention, then we say that $p_{\tau,t}$ is 0.
    
    We use the relative frequencies to construct several other measures of attention. First, we define the proportion of days that a name received attention as the proportion of days since the date of death that a name was within the top million 2-grams. Second, we say that the \emph{peak attention} given to a name is the highest relative frequency with which it was used, i.e. the maximum of $p_{\tau, t}$ across all days $t$. We use the peak attention to construct the \emph{normalized attention}
    \begin{equation}
        \widehat{p_{\tau,t}}
            =
                \frac{
                    p_{\tau, t}
                }{
                    \max_t p_{\tau, t}
                }.
    \end{equation}
    The normalized attention is a value between 0 and 1, where 0 indicates a name was not used within the top million 2-grams for a day, and 1 indicates the day that the most relative attention was given to a name since the date of death. 
    
    Similarly, we define two further measures of attention by measuring the rank of how frequently each name was used on each day $r_{\tau,t}$, and its peak rank, the minimum of $r_{\tau, t}$ across all days $t$ (since lower values of $r_{\tau, t}$ indicate a higher rank). The rank and frequency metrics give related, but different measures of collective attention. The frequency-based measures of attention give a sense of how much a name was used relative to all the other times it was used on Twitter. The rank-based measures give us way of interpreting how high or low that attention was compared to all other 2-grams.
        
    We measure the amplification of each name by distinguishing between how often they were used in originally authored tweets (OT) and retweets (RT), where we include the novel part of quote retweets among originally authored tweets. We operationalize amplification as the ratio between the two frequencies for a given name,
    \begin{equation}
        R_{\tau,t} = \frac{
                      f^{(\text{RT})}_{\tau,t}
                  }{
                    f^{(\text{OT})}_{\tau,t}
                  }.
    \end{equation}
    The amplification measure $R_{\tau,t}$ is 1 when a name is used equally often in originally authored tweets and retweets. If $R_{\tau,t}$ is greater than 1, then the name is amplified via retweets more often than it is written itself, and vice versa if $R_{\tau,t}$ is less than 1. Note, this operationalization of amplification differs from one of the most standard approaches to measuring amplification, which is simply counting retweets. By comparing to how much a name is being written in originally authored tweets, we establish a baseline that allows us to discern to what extent the amplification $R_{\tau,t}$ is more or less than what we would expect given how much that name is being spoken about on Twitter. When a name does not receive measurable attention---it does not rank within the top million 2-grams---we say that $R_{\tau,t}$ is 1, since it was not measurably retweeted more nor less than the inattention it received.

    It is important to note that retweets have become increasingly common over time in English language tweets \cite{alshaabi2021growing}. This means that if we look at $R_{\tau,t}$ over longer time windows, as we do here, names that appear later in our study frame will appear to have been more likely to be retweeted. The long-term increase in the rate of retweets is confounded by other factors though, such as changes to Twitter's design and algorithmic curation.
    To account for this, we define the \emph{relative social amplification} \cite{alshaabi2021growing,alshaabi2020storywrangler} as
    \begin{equation}
        R^{\text{rel}}_{\tau,t}
            =   
                \frac{
                    R_{\tau,t}
                }{
                    \sum_{w' \in W_t} 
                        f_{w',t}^{(\text{RT})}
                    /
                    \sum_{w' \in W_t}
                        f_{w',t}^{(\text{OT})}
                }.
    \end{equation}
    That is, we normalize the amplification of a name by the ratio of how much all English language is used in retweets versus originally authored tweets on any given day. When retweets increase generally, the sum of all language used in retweets will increase, and so the denominator will increase as a whole. This adjusts for how $R_{\tau,t}$ may itself increase because of such increases in retweets. So, unlike $R_{\tau,t}$, the relative social amplification $R^{\text{rel}}_{\tau,t}$ is comparable over wide time frames on Twitter, allowing us to compare the amplification of names to themselves and one another over time.

% ========================================================
% ========================================================
% ======================= Results ========================
% ========================================================
% ========================================================
\section{Results}
\label{sec:results}

% -----------------------------------------------------------------
% -------------------- Results: George Floyd ----------------------
% -----------------------------------------------------------------
    \subsection{Unprecedented Attention in the Wake of George Floyd's Death}
    \label{sec:spike}

    \begin{table*}[!htp]
        \centering
        \begin{tabular}{c|c|c|c|c}
            Spike Period 
            & \shortstack{Number of Names with\\Increased Attention}
            & \shortstack{Perc. of Names with No \\Attention 30 Days Before}
            & \shortstack{Avg. Diff. in Rel. Freq.\\Spike - Before}
            %& $p$-value
            & \shortstack{Avg. Diff. in Rel. Freq.\\After - Before} \\
            %& $p$-value \\%& $\mu_3 - \mu_2$ \\
            \hline
            \hline
            Nov. 24--Dec. 8, 2014
                & 81
                & 71.6\%
                & 3.82e-08 
                %& 0.106
                & 3.37e-09 \\
                %& 0.225\\% & -3.49e-08 \\
            \hline
            Jul. 13--Jul. 26, 2015
                & 36
                & 55.5\%
                & 5.48e-10 
                %& 0.626
                & 4.49e-10 \\
                %& 0.674\\% & -9.96e-11 \\
            \hline
            Jul. 5--Jul. 13, 2016
                & 68
                & 66.1\%
                & 6.83e-09 
               % & 0.061
                & \textbf{2.76e-09*}\\
                %& 0.021\\%& -4.08e-09 \\
            \hline
            Aug. 12--Aug. 22, 2017
                & 34
                & 32.3\%
                & -1.40e-10 
                %& 0.950
                & -7.88e-10 \\
                %& 0.354\\% & -6.49e-10 \\
            \hline
            May 25--Jun. 6, 2020
                & 186
                & 72.3\%
                & \textbf{5.14e-08**}
                %& 0.004
                & 3.43e-08
                %& 0.119\\ %& -1.71e-08 \\
        \end{tabular}
        \caption{
        \textbf{Resurgent attention to past victims of police violence following George Floyd's death and other periods of interest to Black Lives Matter}. We consider four periods of spikes in attention relevant to \#BlackLivesMatter: November 24th--December 8th, 2014 (Deaths and non-indictments in the cases of Michael Brown, Tamir Rice, and Eric Garner), July 13th--July 26th, 2015 (death of Sandra Bland), July 5th--July 13th, 2016 (deaths of Philando Castile and Alton Sterling), August 12th--August 22nd, 2017 (``Unite the Right'' Charlottesville rally), and May 25th--June 6th, 2020 (death of George Floyd). The number of names that received increased attention during a spike period is reported, as well as the percentage of those that had not received any measurable attention in the 30 days prior to the spike. The average change in average relative frequency is calculated for the difference between 30 days before the spike period and during it, and 30 days before and after it. Statistical significance is indicated by * for $\alpha=0.05$ and ** for $\alpha=0.01$.}
        \label{tab:resurgence}
    \end{table*}

    We start by characterizing the impact of George Floyd's death on Twitter in late May 2020, establishing the ways in which it was a pivotal moment in the history of both Black Lives Matter and Twitter more broadly. George Floyd's death on May 25th, 2020 was followed by a massive increase in tweet volume lasting until about June 7, with a peak on June 2nd (see middle panel of Figure~\ref{fig:spike}). 
    Approximating from the 10\% Decahose random sample of tweets, we estimate that an average of 197 million tweets were sent per day during that period, with up to about 219 million tweets per day between May 31 and June 2, 2020. 
    Those latter three days constitute the 2nd, 3rd, and 4th days with the most tweets overall compared to all other days in our Decahose sample, dating back to 2009. Further, they are the top three days in Twitter history in terms of the number of retweets that were sent.
    For comparison, we estimate about 157 million tweets were authored per day from May 1st to May 26th, 2020.

    As shown in Figure~\ref{fig:spike} and detailed further in Table~\ref{tab:spike}, the increase in tweet volume was driven almost entirely by retweets.
    The average number of originally authored tweets per day increased from 69.7 million between May 1st and May 26th to 71.2 million between May 27th and June 7th---an increase of only 2.1\%. 
    Meanwhile, the average number of retweets per day rose from 87.5 million between May 1st and May 26th to 126 million between May 27th and June 7th, yielding a dramatic 44\% increase.
    This spike period is characterized by a simultaneous spike in the discussion of Black Lives Matter (top panel of Figure~\ref{fig:spike}).

    \begin{figure*}[!htp]
      \centering	
        \includegraphics[width=\textwidth]{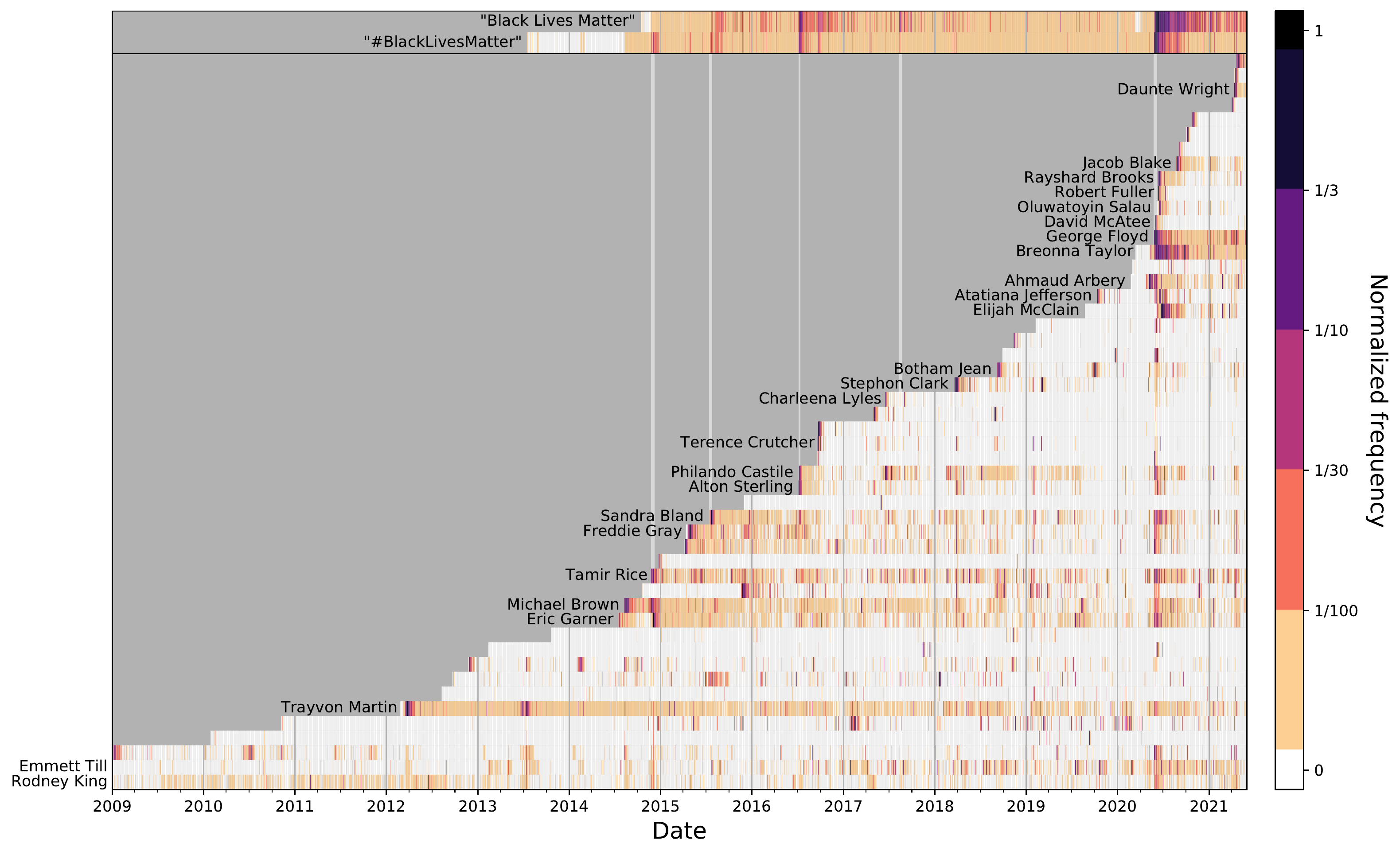}  
      \caption{
        \textbf{Attention towards emblematic names of Black Lives Matter.} 
        \textbf{Top)} Heatmap time series of the normalized relative frequency of ``Black Lives Matter" and ``\#BlackLivesMatter".
        \textbf{Main)} Heatmap time series displaying the normalized attention $\widehat{\ngramprop}$ over time for the top 50 names in the combined database by peak rank. The top 25 names are labeled. Darker colors indicate more attention, where the normalized attention is 1 (black) on the date of peak attention.
        Dark bands around June 2020 are observed across many of the rows, indicating a resurgence in attention for many of the emblematic names.
      }
      \label{fig:heatmap}
    \end{figure*}
    
    The unprecedented increase in tweet volume during the spike period resulted in ``George Floyd'' being the 7th most used 2-gram on Twitter on May 29th, 2020. To emphasize how exceptional this rank is ~\cite{dodds2019fame}, we note that only the 2-grams ``!!'', ``of the'', ``??'', ``in the'', ``, and'', and ``to be'' appeared more frequently than ``George Floyd'', with ``to the'', ``on the'', and ``. I'' rounding out the top ten.
    Only a few dozen other proper names have reached similar levels of attention on Twitter. Others include ``Muhammad Ali'' who reached a rank of 2 on June 4th, 2016, the day after his death, and ``Donald Trump'' who reached a rank of 6 on November 9th, 2016 the day following his election as president of the United States.
    Among those who have reached such stratospheric levels of attention, George Floyd is the only one who was unknown to the public earlier the same week. 
    %His name has been mentioned every day on Twitter since his death. 

    The increased volume in tweets corresponded with a decreased expression of happiness on Twitter (see bottom panel of Figure~\ref{fig:spike}). Based on our 12-year Decahose sample, May 31st, 2020, six days after George Floyd's death, was the day with the least happiness expressed in Twitter's history. The average happiness expressed on that day was 5.628, while the average happiness expressed in the preceding seven-day period was 5.803. For reference, the average happiness expressed from 2009 to 2021 is 6.012 with a standard deviation of 0.062. This drop in happiness was unique not only in its intensity, but also its duration. 
    Prior to the COVID-19 pandemic, nearly all decreases in happiness caused by tragedies---like other police violence events, mass shootings, and celebrity deaths---have lasted only about a single day before returning to ``normal'' levels\footnote{See 
    \url{http://hedonometer.org}} \cite{Sharkeye2100846118}.
    It is difficult to gauge a ``normal" level of happiness expressed on Twitter, but the happiness expressed in June 2020 did not reach a level of 5.9 or greater again until June 21st (Father's Day). 
    The only other period with such an evidently long recovery was during March and April 2020 when the coronavirus pandemic dramatically altered daily life in the United States.

    %HW: I need to say that the duration of this decrease is much longer than past police violence events
    During the slow regression back toward ``normal'' happiness on Twitter, discussion shifted from the event of George Floyd's death to the protests following it. 
    Examining the individual words that contributed most to the decrease in expressed happiness, we find ``murder,'' ``killed,'' and ``death'' had strong contributions shortly after his death. Words such as ``violence,'' ``protest,'' and ``terrorist'' then began to show stronger contributions in the following days, suggesting that the collective conversation began to move away from Floyd's death and toward the protests themselves. Supplementary Figures~\ref{fig:saddestday}--\ref{fig:wordshiftjune7} fully summarize the words that contributed the most to the decreased expression of happiness.

    \begin{figure*}
      \centering	
        \includegraphics[scale=0.45]{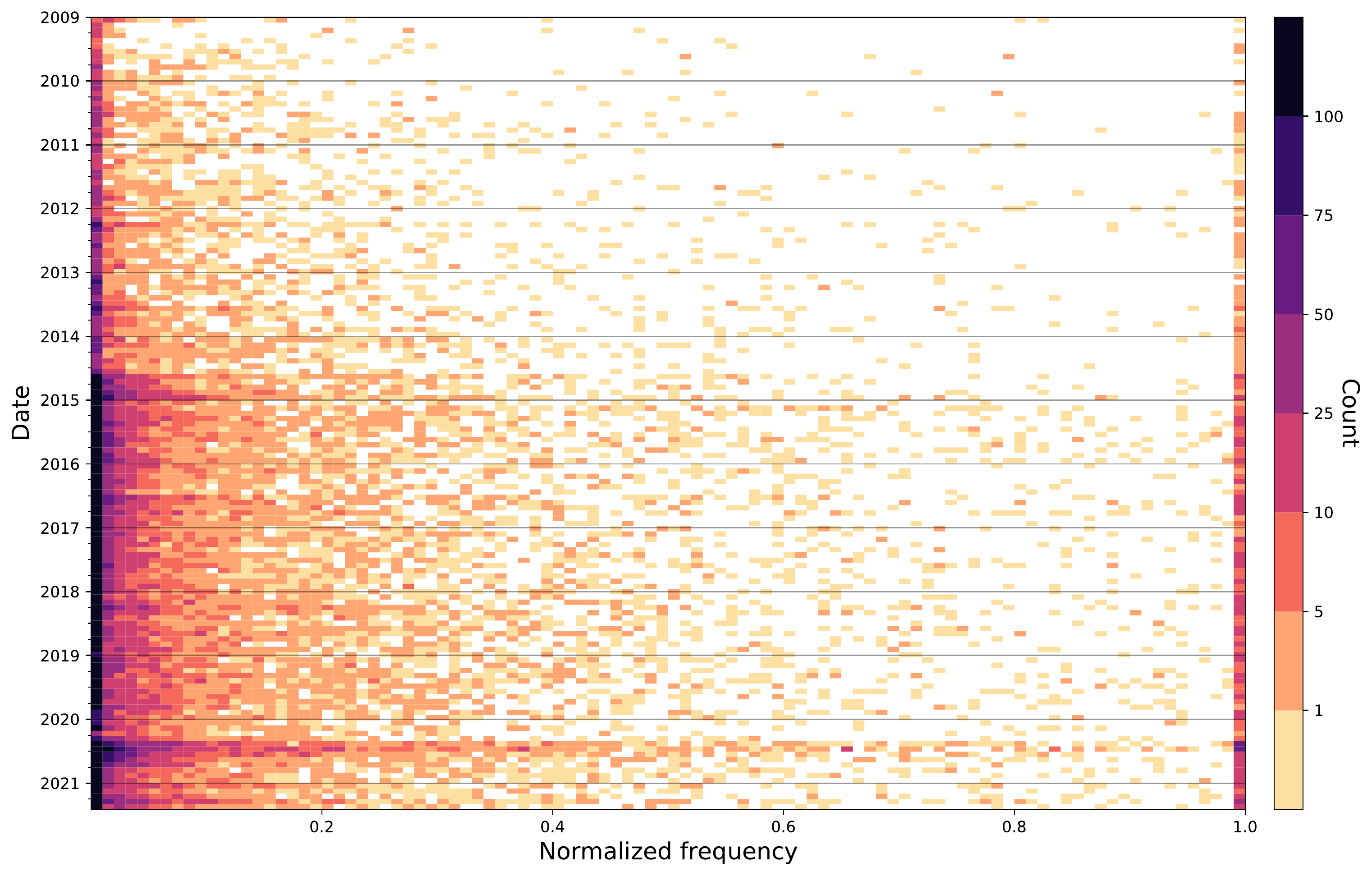}  
      \caption{
        \textbf{Collective attention toward names over time.} 
        Histograms over time showing the amount of normalized attention $\widehat{\ngramprop}$ given to Black victims of fatal police violence from January 1, 2009 to May 31, 2021 (intervals of about 30 days). Each cell indicates the number of names that received that particular amount of attention in that moment of time.
        %The bin size is approximately 30 days in the vertical direction and 0.01 in the horizontal direction. 
        There is a band around May--June 2020 across many different frequencies, indicating a variety of attention for many names during that period.
      }
      \label{fig:2dhist}
    \end{figure*}

% -----------------------------------------------------------------
% -------------------- Results: Resurgence ------------------------
% -----------------------------------------------------------------   
    \subsection{Resurgence of Past Names}
    \label{sec:resurgence}

    In the wake of George Floyd's death, the historic amount of attention was not just given to him nor even just to the subsequent protests. It also extended to past Black victims of fatal police violence. To show this, we measure the extent to which their names received heightened attention during and shortly after the spike in Twitter activity from May 25th to June 7th. For each name, we calculate the average relative frequency with which it was used thirty days before the spike. We then compare that to its average relative frequency during the spike, and thirty days following the spike. For reference, we compare the average change in relative frequency during the George Floyd spike to the average change during three other important spikes in Black Lives Matter history: the emergence of \#BlackLivesMatter around the deaths of Michael Brown, Tamir Rice, and Eric Garner (November 24th--December 8th, 2014), the introduction of \#SayHerName and the death of Sandra Bland (July 13th--July 26th, 2015), and the deaths of Philando Castile and Alton Sterling (July 5th--July 13th, 2016). We also compare to the ``Unite the Right'' rally in Charlottesville, Virginia (August 12th--August 22nd, 2017), during which time use of the hashtag \#BlackLivesMatter spiked but not due to a police-involved death.

    More attention was given to Black victims of fatal police violence during the spike following George Floyd's death than any of the other spikes, and that attention persisted longer following the spike (see Table~\ref{tab:resurgence}). The average change in average relative frequency is an order of magnitude larger during the George Floyd spike than the ones following the deaths of Sandra Bland, Philando Castile, and Alton Sterling, and 34\% larger than when \#BlackLivesMatter first saw widespread use in late 2014.  Relative to the pre-spike periods, while the average change in attention to names decayed by an order of magnitude after the initial spike of \#BlackLivesMatter and by 59\% following the deaths of Philando Castile and Alton Sterling, it only declined by 33\% as the spike following George Floyd's death subsided. As expected, use of Black victims' names did not increase significantly during the ``Unite the Right'' rally, as police-involved deaths were not the topic of concern. We find these results are robust even if we vary the pre- and post-spike time windows to be anywhere from 7 to 90 days long (see Supplementary Table~\ref{tab:resurgence90}).

    Further, the increase in attention to victims of fatal police violence was not just caused by attention to names that are most emblematic of the Black Lives Matter movement (e.g. Michael Brown, Sandra Bland). The names of 186 victims were used more during the spike following George Floyd's death than they were in the thirty days before. This is more than twice as many victims that were mentioned in any of the other spike periods that we look at here, and over 70\% of those names had not been mentioned at all in the month proceeding Floyd's death. In Supplementary Table~\ref{tab:namelist1}, we display the names of all those who received more attention on average during the spike than prior to it.

         \begin{figure*}
      \centering	
        \includegraphics[width=0.8\textwidth]{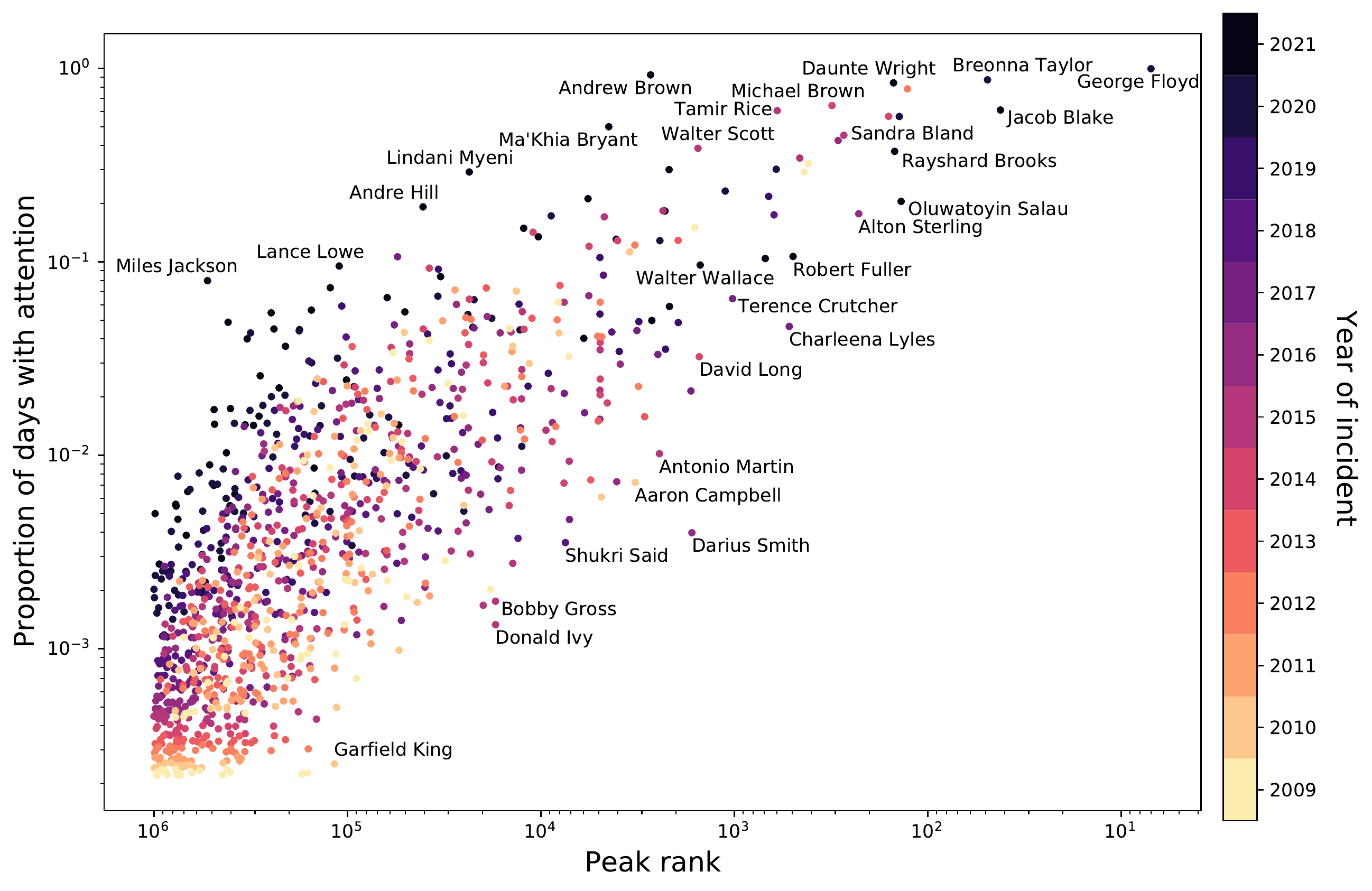}  
      \caption{
        \textbf{Proportion of days with attention versus the peak rank of names}. 
        %A scatterplot of the names in the combined database displaying each name's peak rank (x-axis) and the proportion of days since death (or event) on which it reached the top million in rank (y-axis). 
        Each point is colored by the year of the event or death. 
        There is a moderately strong trend---as peak rank increases, the proportion of days mentioned 
        %(among the top million 2-grams) 
        increases.
        Names with a relatively high proportion of days with attention and a relatively high peak rank tend to be more recent deaths. 
        %See Appendix Figs.~\ref{fig:namerankmale} and~\ref{fig:namerankfemale} for the same figure separated by gender.
      }
      \label{fig:namerank}
    \end{figure*}

    This is not to say that prior surges in attention to Black Lives Matter have not been effective at highlighting instances of police violence, nor that they did not have any lasting impact. Rather, the increased attention following George Floyd's death should be understood as the culmination of over 6 years of movement building since \#BlackLivesMatter first gained traction following the death of Michael Brown in 2014.

% -----------------------------------------------------------------------
% -------------------- Results: Long-term dynamics ----------------------
% -----------------------------------------------------------------------
    \subsection{Long-Term Trends of Attention and Amplification}
    \label{sec:namerank}
    
    To put May and June 2020's exceptional levels of attention to police violence in context and understand how that moment built off the foundations laid by the Black Lives Matter movement and other activists, we step back and view how attention has been given to police-involved deaths over the past 12 years on Twitter. We take this view from two different perspectives: that of attention and that of amplification. Together, these paint a detailed picture of how individuals on Twitter have reiterated the names of Black victims of police violence.

    \begin{figure*}
      \centering	
        \includegraphics[width=0.95\textwidth]{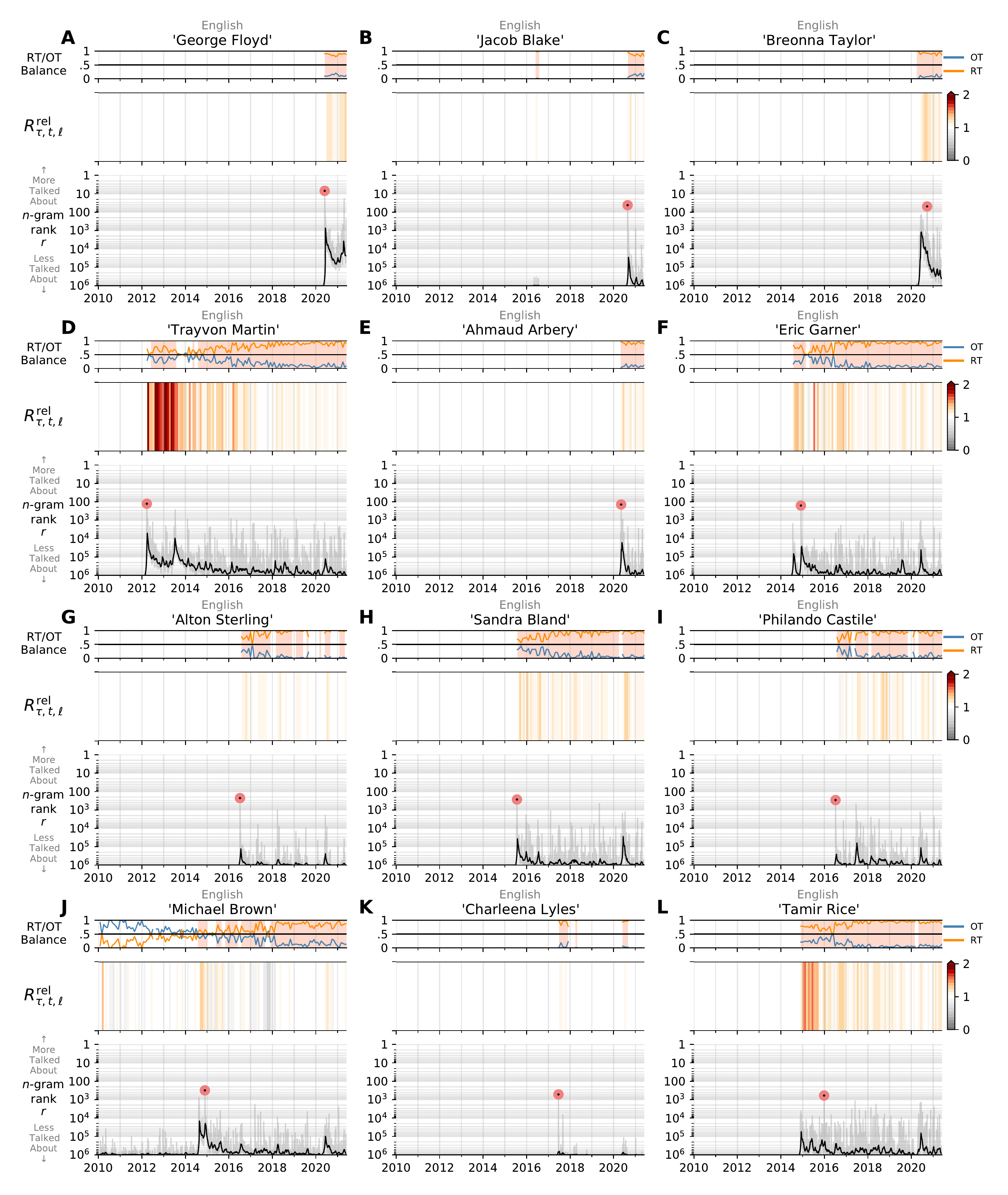}  
      \caption{
        \textbf{Amplification of emblematic names of Black Lives Matter.} Each panel visualizes amplification and attention in three ways: \textbf{Top)} Time series of the proportion of mentions of a name that were from originally authored tweets (blue) and retweets (orange). \textbf{Middle)} Heatmap time series of the relative social amplification $R_{\tau,t}^{\text{rel}}$ of a name relative to all English language. When the relative social amplification is greater than 1, that period is highlighted in orange in the top panel as well; this means that the name is amplified (retweeted) relatively more than other English $n$-grams of the same length. \textbf{Bottom)} Time series of a name's daily rank compared to other $n$-grams of the same length. The 7-day rolling average is shown in bold, and the day of peak rank is indicated with a pink dot. See Supplementary Figure~\ref{fig:namecontagiogramssupp} for other selected names.
      }
      \label{fig:namecontagiogramsmain}
    \end{figure*}

    % -----------------------------------------------------------------
    % -------------------- Results: Attention -------------------------
    % ----------------------------------------------------------------- 
    \subsubsection{Collective Attention}

    In Figure~\ref{fig:heatmap}, we show how emblematic names of the Black Lives Matter movement have been used on Twitter across the past decade. Although \#BlackLivesMatter did not gain traction until 2014, it was created in 2012 following the death of Trayvon Martin. Even without the hashtag though, we see that his name has been consistently used since his death. With the deaths of Eric Garner, Michael Brown, Tamir Rice, Freddie Gray, and Sandra Bland, we see increases in attention to different prominent victims of fatal police violence, and that the attention to those names was sustained as the movement gained its ground. While there were a number of high-profile incidents over the latter half of the decade, there is a stark vertical band around May and June 2020 following George Floyd and Breonna Taylor's deaths in which many of the most recognizable victims of police violence all received high levels of attention. This is one visual indicator that reaffirms the resurgence in names in the wake of George Floyd's death.

    However, as we have discussed, the most high-profile incidents of fatal police violence are just a fraction of all incidents. In Figure~\ref{fig:2dhist}, we show the spectrum of how much attention has been given to all Black victims of police violence over time. Again, we see a notable increase in attention around late 2014, as the hashtag \#BlackLivesMatter gained traction. Compared to pre-2014, the amount of attention given to names has been consistently higher, as indicated by the increase in dark cells. The strong band of color around May and June 2020 is again visible, showing that not only did the most prominent names gain increased attention (note the band on the right-hand side of the plot), but also many other names did as well, at various frequencies. Note, there is a dark band on the left-hand side of the plot as well: while the names of many victims have been mentioned on Twitter and increasingly so over time, there are also many others do not see such widespread attention.

    In Figure~\ref{fig:namerank}, we unpack the relationship between the number of days that different names have received attention, and how widely they have been discussed, as measured by their peak rank compared to all other 2-grams on Twitter for a given day. Among those with the most and widest attention, we see the emblematic Black Lives Matter names, like Breonna Taylor, Walter Scott, Alton Sterling, and Sandra Bland. The plot also reveals the number of names that do not receive those heights of attention. Most police-involved deaths of Black victims generally have not consistently received attention, nor have they they had high peak ranks. Given that most incidents receive their peak attention within a week of their occurrence (see Supplementary Figures~\ref{fig:attentiondecay}--\ref{fig:peakdelayhist}), this makes it difficult for them to ever receive high attention if they do not so immediately. This is particularly clear among the incidents that occurred around 2010, which we see generally have lower peak ranks, and lower proportions of days with attention. There are notable exceptions in either direction of the proportion of days with attention and the peak rank. For example, some (e.g. Darius Smith) received a relatively high amount of attention (i.e. high peak rank) at one point, but have not been given attention for many days in general. Others (e.g. Patrick Warren, Sincere Pierce) never reached a relatively high level of discussion, but have still been more consistently discussed since their deaths than others. In Supplementary Figures~\ref{fig:namerankmale} and \ref{fig:namerankfemale}, we further detail the relationship between proportion of days with attention and peak rank by gender. Overall, the figures demonstrate the range of attention that has been given to Black victims of police violence, and how that attention extends beyond the most emblematic names of the Black Lives Matter movement.

    % -----------------------------------------------------------------
    % -------------------- Results: Amplification  --------------------
    % ----------------------------------------------------------------- 
    \subsubsection{Relative Social Amplification}

    \begin{figure*}
      \centering
        \includegraphics[scale=0.5]{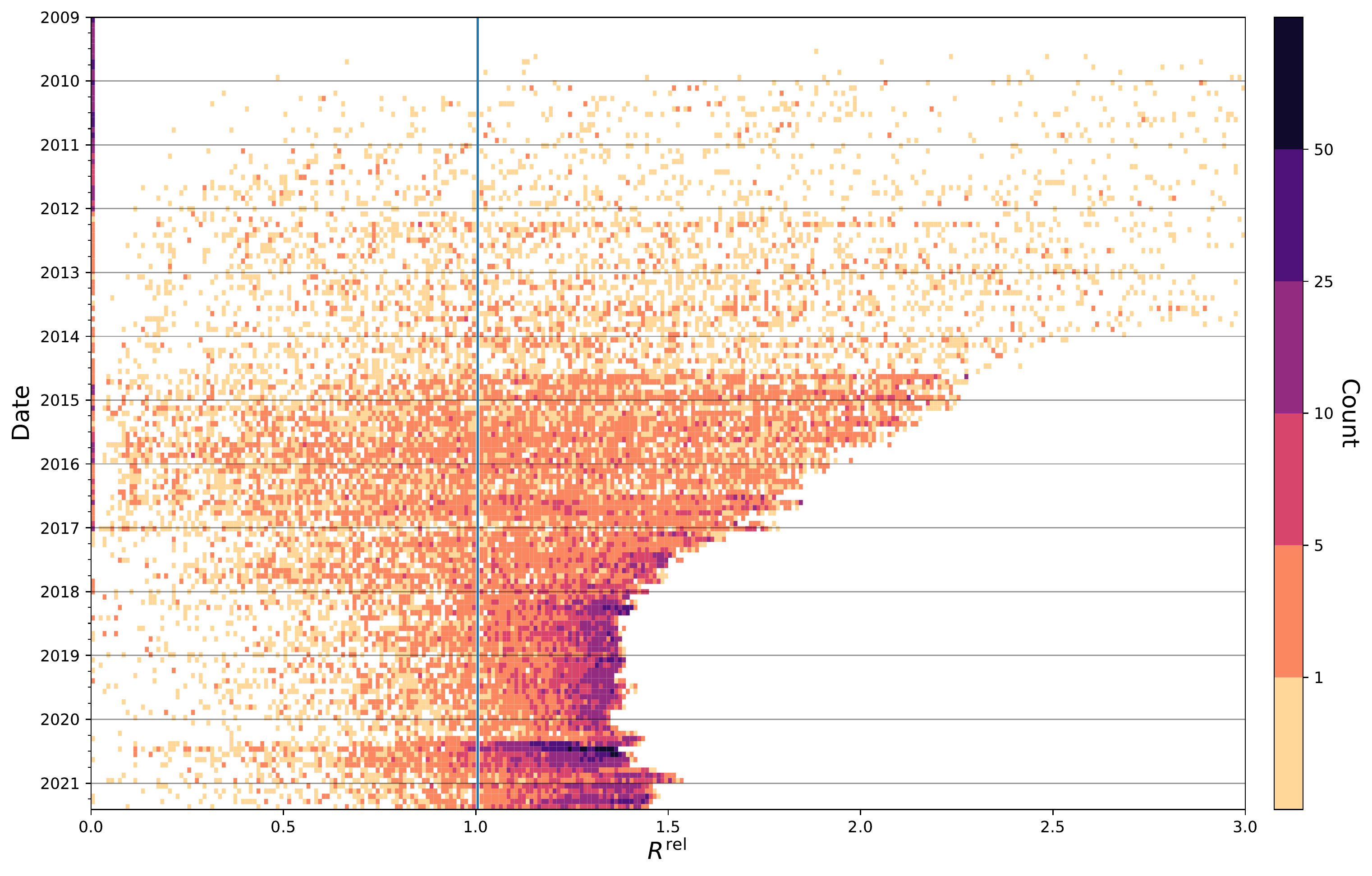}
      \caption{
        \textbf{Relative social amplification of names over time.} 
        Histograms over time showing the relative social amplification $R_{\tau,t}^{\text{rel}}$ given to Black victims of fatal police violence from January 1, 2009 to May 31, 2021 (intervals of about 30 days). Each cell indicates the number of names that received that particular level of relative social amplification in that moment of time. 
        Names increasingly have a relative social amplification higher than 1, meaning they are amplified (retweeted) relatively more often than other English 2-grams.
        Around May and June 2020, more names take on a high $R_{\tau,t}^{\text{rel}}$ of about 1.4, indicating high social amplification for many names during that period.
        The maximum relative social amplification for names decreases over time because the proportion of all English tweets that are retweets increases over time \cite{alshaabi2021growing}.
      }
      \label{fig:ratiohist}
    \end{figure*}

    As with attention, amplification of the names of Black victims of fatal police violence has also varied over the past decade on Twitter. In Figure~\ref{fig:namecontagiogramsmain}, we show the different trajectories of how some of the most emblematic names of the Black Lives Matter movement have been amplified over time. For each name, we measure both the proportion of tweets containing the name that were originally authored tweets and retweets, and how amplification of that name changed relative to English language as a whole. Like attention, we can see that the amplification of Trayvon Martin clearly predates the widespread use of \#BlackLivesMatter, and his name has been consistently amplified since his death. A feature of some of the time series---including his, but also others such as Michael Brown, Eric Garner, and George Floyd---are two distinct peaks in attention with regard to their rank $r_{\tau,t}$. The dual peaks for these cases correspond to the initial attention given to the victim's death, and the later (non-)indictment or trial. For some, the peak rank of their name on Twitter is higher at the time of legal judgement---\#BlackLivesMatter, for example, only gained traction after the non-indictment of Darren Wilson for the death of Michael Brown, and Michael Brown's name was used more then than at the time of his death. Further examples of these amplification trajectories are shown in Supplementary Figure~\ref{fig:namecontagiogramssupp}.
    
    \begin{figure*}
      \centering	
        \includegraphics[width=0.95\textwidth]{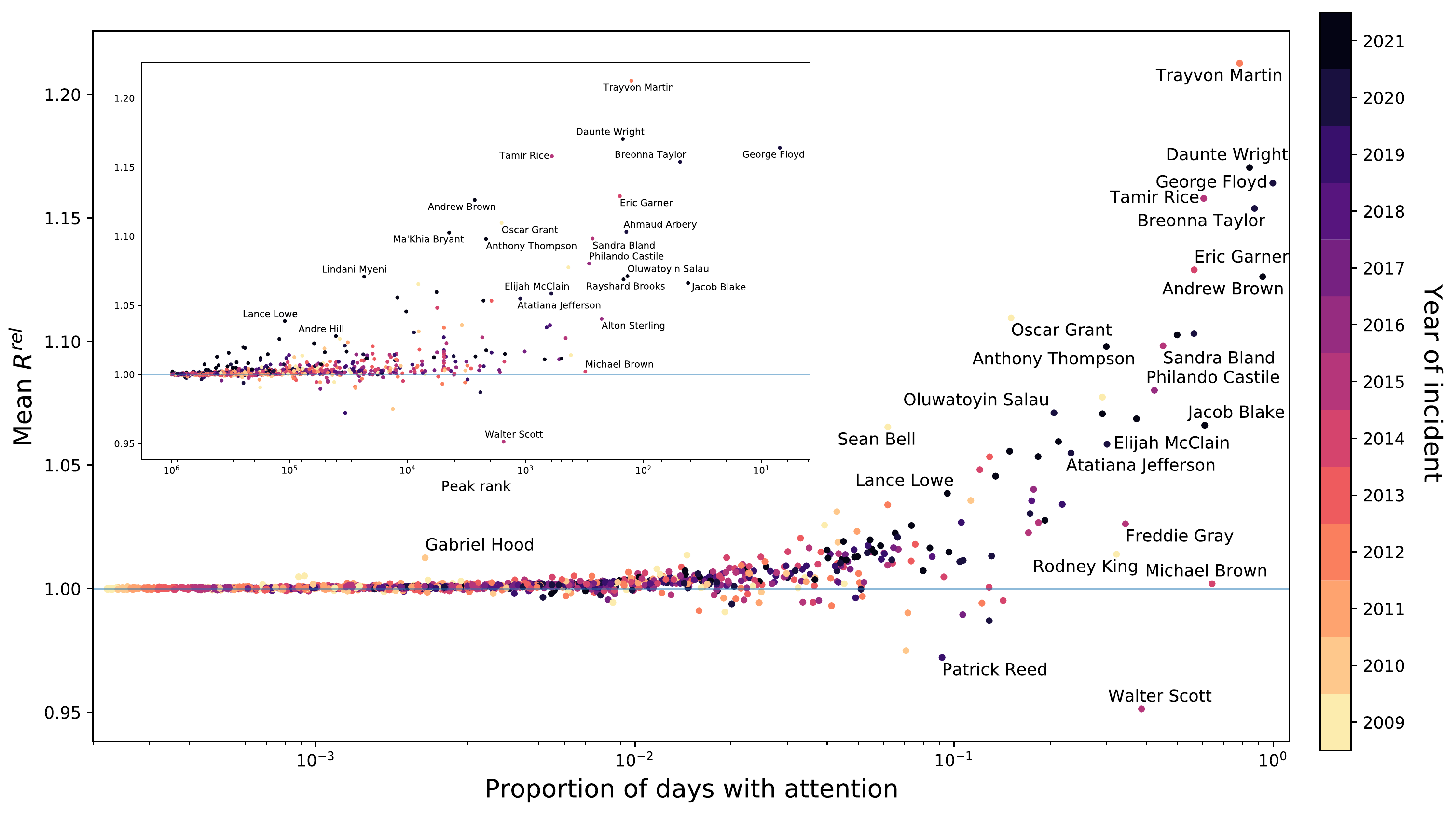}
      \caption{
        \textbf{Mean relative social amplification versus proportion of days with measurable attention and peak rank.} 
        %The proportion of days with measurable attention is defined as the fraction of days since the date of death where the rank is in the top million. The y-axis in both the main and inset figure is the mean relative social amplification $R_{\tau,t}^{\text{rel}}$. A value greater than 1 means that the name used in retweets relatively more often than in organic tweets compared to the English language as a whole, while a value less than 1 means the opposite. A horizontal blue line marks this threshold. 
        The mean relative social amplification $R_{\tau,t}^{\text{rel}}$ of most names is close to 1, but as the peak rank and proportion of days with measurable attention increase, the range of relative social amplification values widens. Each point is colored by the year of the event or death. 
      }
      \label{fig:namerank2}
    \end{figure*}

    As we see by these examples, amplification ebbs and flows, even for the most emblematic names. Again, given that most police-involved deaths are not those that go viral through \#BlackLivesMatter, we look in Figure~\ref{fig:ratiohist} at the amplification of the names of those who received measurable attention. In late 2014, we see more amplification of a wider variety of names compared to the pre-\#BlackLivesMatter era. We see this amplification consolidate in two ways. First, over time, more names receive more relative social amplification. In particular, since 2014, most names of Black police violence victims that receive measurable attention on Twitter have a relative social amplification $R_{\tau,t}^{\text{rel}}$ greater than 1, meaning that they are amplified (retweeted) more than other English 2-grams. In other words, the names of Black victims of fatal police violence receive more amplification than we would expect compared to English language amplification generally. We also see a consolidation of amplification: names generally are no longer amplified more than about 1.5 times more than other English $n$-grams, as compared to the early 2010s, during which time the relative social amplification could be upwards of three times higher (or more, see Supplementary Figures ~\ref{fig:meanratiohist} and \ref{fig:maxratiohist}). This is likely a result of the increase in the proportion of tweets that are retweets in English in general \cite{alshaabi2021growing}. Although the relative social amplification may not reach as drastic of heights, whether that be due to platform design changes or alterations to how Twitter curates timelines, there is a much higher number of names that receive amplification, and---if a name receives measurable attention on Twitter---most of the time that amplification exceeds what we would expect otherwise.

    Finally, we compare the average relative social amplification of each name $R_{\tau,t}^{\text{rel}}$ to its peak rank and the proportion of days it received measurable attention (see Figue~\ref{fig:namerank2}). On average, the relative social amplification of most names is 1, meaning that they are amplified as commonly as we would expect relative to how much all other English language is amplified. We see though that names with the highest proportions of days with attention and highest peak ranks are those that receive the most amplification on average. This makes the barrier of entry into the mainstream public sphere \cite{jackson2015hijacking} clear: it is difficult for a name reach a high level of visibility (high peak rank) or consistently receive attention without a consistently high level of relative social amplification. That is, this demonstrates a link between attention and amplification, where it is likely the case that it is both difficult to receive widespread attention without high levels of amplification, and difficult to receive amplification without high levels attention. This likely contributes to why most Black victims of police violence never receive measurable attention on Twitter.

% ===========================================================
% ===========================================================
% ======================= Discussion ========================
% ===========================================================
% ===========================================================
\section{Discussion}
\label{sec:discussion}
    
    This study contributes a bird’s-eye view of how the Black Lives Matter movement has brought Black victims of police violence to the forefront by invoking their names, connecting new instances of police violence with past ones. In particular, we have shown that there was an exceptional resurgence in attention to past victims of police violence following the death of George Floyd. His death itself had an unprecedented impact on Twitter: it prompted the 2nd, 3rd, and 4th most tweets to be sent per day in Twitter’s history, where those three days have the most retweets of any previous day on Twitter. The historic levels of tweeting were matched with a historic dip in happiness expressed on the platform: over 12 years of Twitter data, we have not seen happiness drop as low as it did on May 31st, six days after Floyd’s death and in the midst of the subsequent protests. The attention to his death though came with exceptional levels of attention to \emph{other} past victims as well, particularly as compared to prior spikes in attention to the Black Lives Matter movement. Understanding that those levels of attention built off prior surges in attention to \#BlackLivesMatter, we put the surge of attention following Floyd’s death within the context of how the names of Black victims of police violence have been given attention over the past 12 years on Twitter.

    The long history of systemic racism and police brutality invites the act of memorialization, a central tactic of the Black Lives Matter movement. Victims' names are used to honor and draw attention to specific individuals and, especially when named in combination, to connect individual deaths with the history of state violence against Black Americans. In this way, names become signifiers of broader injustices, with particular combinations evoking various dimensions of injustice to be addressed. For example, pairing Emmett Till and Trayvon Martin might signify state violence against Black youth, while pairing Sandra Bland and Breonna Taylor might signify state violence against Black women (and its corresponding erasure in media coverage). The evocative nature of names as signifiers allows Black Lives Matter activists to tell complex, intersectional stories about police violence in the limited space of a tweet.

    Our analyses bring up many directions for further understanding how signifiers are used in the Black Lives Matter movement and other instances of online activism. Here, we have focused on comprehensively characterizing the different ways in which names have been given attention on Twitter. This invites further investigation of what factors---like a victim’s gender or age, or whether there is video footage of an event---determine the amount of attention that a name receives. In doing so, it is also important to move from measuring aggregate collective attention to measuring how individual users and groups of users contribute to attention and amplification. How we understand the attention given to a particular victim may differ depending on whether that attention results from a small number of viral tweets or many original tweets, or whether it comes from liberal or conservative accounts. These questions are naturally related to how Twitter users are connected to one another online, and how those networks affect the extent to which different names are amplified.

    Of course, there are also questions of how the online attention relates to offline impact. Data on offline racial justice protests and tweet volume around particular victims could be used to quantify how social media is used to organize protests, donations, and petitions for Black Lives Matter and other social movements. Because we started with an extensive list of victims of police violence, rather than focusing just on cases which received national levels of attention, this would contribute to our understanding of how online discussion can translate to offline activism without selecting on the most visible instances of online activism. Such attention, online and offline, may be affected by events other than death: birthdays, anniversaries, and trials may also uniquely contribute to resurgences in attention to those who have been killed by police violence. More broadly, the methods used here to analyze attention towards police violence victims could be applied to study other political figures, historical events, and social justice movements.

    Our study rests on the Fatal Encounters database, a third-party repository for recording police-involved deaths. This is because there is a glaring lack of government-maintained police violence databases, and there is a clear need for better tracking and reporting of how police use force in the field. Merging the database with the social media data required us to make choices about inclusion and exclusion. Specifically, some of these choices involved attributing mentions of a duplicate name to just the earliest instance of that name, and excluding those who had names that received attention prior to their death, some of which were shared with celebrity figures. More sophisticated methods for determining the context of how a name is being used in a tweet would allow us to more inclusively study victims of police violence.

    Twitter is only one platform on which these names can be given attention. Overall, it is not representative of the general population~\cite{mellon2017twitter, greenwood2016social,malik2015population, zheng2018survey}, though it is worth emphasizing it still plays a notable political role because many journalists, political figures, and others with significant offline platforms heavily use the social media site. Because we only use the Decahose feed for our analysis though, we cannot speak to the extent to which different demographics gave attention to victims of police violence and amplified their names. It would be beneficial to use a panel-based approach~\cite{shugars2021pandemics} to understand differences in how different racial, political, and other demographic groups engage with instances of police violence. Looking beyond Twitter though, it would be valuable to understand the ways in which those on different platforms, like Facebook, Reddit, and TikTok, give attention to and amplify the names of Black victims of police violence.

% ===========================================================
% ===========================================================
% ======================= Conclusion ========================
% ===========================================================
% ===========================================================
\section{Conclusion}
\label{sec:conclusion}
    Remembering Black victims of police violence as people, first and foremost, is key to understanding the effects of racism and enacting policy that addresses it. Here, we have shown that since the growth of the Black Lives Matter movement in 2014, people have regularly and increasingly memorialized those victims by saying their names. While a substantial number of Black people have been recognized as victims of fatal police violence, we emphasize that the majority of Black victims still receive little attention online. There are many reasons that may be the case. Our study presents a moment to reflect on how we give attention to and amplify Black victims of police violence, and what inequities may still exist in how we do so. As the Black Lives Matter movement enters its seventh year of campaigning for racial justice, we all have an opportunity to ask how we can best memorialize and honor victims of police violence by saying their names.

\acknowledgments
The authors are grateful for financial support from the MassMutual Center of Excellence in Complex Systems \& Data Science, and Google Open Source under the Open-Source Complex Ecosystems And Networks (OCEAN) project.
We thank many of our colleagues 
at the Computational Story Lab 
for their feedback on this project.
Computations were performed on the Vermont Advanced Computing Core supported in part by NSF award No. OAC-1827314.

\bibliographystyle{unsrtabbrv}

\clearpage

\onecolumngrid

\newwrite\tempfile
\immediate\openout\tempfile=startsupp.txt
\immediate\write\tempfile{\thepage}
\immediate\closeout\tempfile

\renewcommand{\thesection}{S\arabic{section}}
\renewcommand{\thesubsection}{S\arabic{section}.\arabic{subsection}}
\renewcommand{\thefigure}{S\arabic{figure}}
\renewcommand{\thetable}{S\arabic{table}}
\setcounter{section}{0}
\setcounter{figure}{0}
\setcounter{table}{0}
\setcounter{footnote}{0} 

\clearpage
\section{Data Preprocessing}
\label{sec:datasetdetails}

    We take all fatalities of Black people in the Fatal Encounters database~\cite{finch2019using}  from January 1, 2009 onward. There are 5,546 names that match these criteria. After removing deaths by suicide and vehicular and pursuit deaths, we are left with 3,897 cases. We parse the name of each Black victim into a 2-gram (two words) by removing periods, quotation marks, commas, and parentheses throughout the name, removing ``Jr.," ``Sr.," and Roman numerals from the end of each name, and taking the first and last word in each name. This is done to reflect common name conventions, although this method misses people who go by nicknames or middle names. In instances of duplicate names, where two or more individuals who suffered fatal police violence share a name, we attribute all the mentions of the name to the earlier of the two incidents. In total, there are 54 duplicate names. Another 95 names are excluded because they received measurable attention in the 10 days prior to their death, i.e. they were among the top million 2-grams in the 10 days prior. (One of those names, Walter Scott, was added back in manually.) We manually remove 12 cases of ambiguous names that are not captured by the previous step. Finally, 14 cases with unknown names are excluded, leaving 3,722 cases from the Fatal Encounters database. We list all of the names that were excluded in Tables~\ref{tab:duplicates}-~\ref{tab:unknowns}.\\

\begin{table*}[!htp]
    \centering
    \begin{scriptsize}
    \begin{tabularx}{\textwidth}{rC|rC|rC}
        \textbf{Name} & \textbf{Date} & \textbf{Name} & \textbf{Date} & \textbf{Name} & \textbf{Date} \\
        \hline\hline
        William Smith & 2009-05-17 & Reginald Moore & 2015-05-03 & Charles Baker & 2017-06-02 \\
        James Hill & 2009-07-23 & Richard Davis & 2015-05-31 & Charles Smith & 2018-01-07 \\
        Jerome Williams & 2011-04-18 & Albert Davis & 2015-07-17 & Michael Ward & 2018-03-12 \\
        Brenda Williams & 2011-04-27 & John Allen & 2015-11-04 & Kenneth Ross & 2018-04-11 \\
        Dominique Smith & 2011-07-30 & Charles Smith & 2016-01-31 & Detandel Pickens & 2018-06-23 \\
        Robert Thompson & 2012-09-01 & Eric Harris & 2016-02-08 & Marcus Smith & 2018-09-08 \\
        James Coleman & 2013-04-24 & Kisha Michael & 2016-02-21 & Ronald Singletary & 2018-09-08 \\
        Antonio Johnson & 2013-07-09 & Marquintan Sandlin & 2016-02-21 & Alonzo Smith & 2018-10-10 \\
        Kenneth Thompson & 2013-08-29 & Christopher Davis & 2016-02-24 & Thomas Johnson & 2019-03-19 \\
        Brandon Smith & 2013-10-13 & Michael Wilson & 2016-05-22 & Gregory Edwards & 2019-09-17 \\
        William Jackson & 2013-12-29 & Kendrick Brown & 2016-08-13 & Anthony Smith & 2019-12-26 \\
        Charles Brown & 2014-04-13 & Robert Brown & 2016-09-07 & Malik Williams & 2019-12-31 \\
        Warren Robinson & 2014-07-05 & Gerald Hall & 2016-12-25 & Richard Davis & 2020-01-30 \\
        Frederick Miller & 2014-08-16 & Joshua Jones & 2017-01-20 & Joshua Johnson & 2020-04-22 \\
        Andre Jones & 2014-08-18 & Alonzo Ashley & 2017-02-11 & Michael Harris & 2020-08-29 \\
        Kevin Davis & 2014-12-29 & Eddie Davis & 2017-03-23 & Anthony Jones & 2020-10-12 \\
        Brandon Jones & 2015-03-19 & Kenneth Johnson & 2017-04-12 & Robert Howard & 2021-01-06 \\
        Eric Harris & 2015-04-02 & Darius Smith & 2017-05-26 & James Alexander & 2021-04-07 \\
    \end{tabularx}
    \end{scriptsize}
    \caption{\emph{Duplicate names in the Fatal Encounters database.} We attribute all mentions of a name to the earliest incident with a victim of that name, indicated by the date in the table.}
    \label{tab:duplicates}
\end{table*}
    
\setlength{\tabcolsep}{2pt}

\begin{table*}[!htp]
    \centering
    \begin{scriptsize}
    \begin{tabularx}{\textwidth}{rC|rC|rC}
        \textbf{Name} & \textbf{Date} & \textbf{Name} & \textbf{Date} & \textbf{Name} & \textbf{Date} \\
        \hline
        \hline
        Lamar Smith & 2009-01-09 & James Brown & 2013-01-25 & Terrance Williams & 2016-11-17 \\
        Robert Johnson & 2009-02-10 & John Harris & 2013-03-14 & George Bush & 2016-11-21 \\
        Kevin Jackson & 2009-04-15 & William Morris & 2013-04-02 & Frank Clark & 2016-11-22 \\
        Kenneth Williams & 2009-05-10 & John Williams & 2013-11-14 & William Boyette & 2017-02-07 \\
        Michael Williams & 2009-06-29 & Jason White & 2013-11-17 & Christopher Carter & 2017-02-19 \\
        Robert Brown & 2009-09-06 & Paul Smith & 2014-01-14 & Don Johnson & 2017-03-23 \\
        Robert Johnson & 2009-09-24 & Dustin Brown & 2014-01-21 & Marcus Williams & 2017-04-01 \\
        Erik Johnson & 2009-10-27 & David Robinson & 2014-03-10 & David Jones & 2017-06-08 \\
        Maurice Clemmons & 2009-12-01 & Gary Smith & 2014-05-11 & Paul Jones & 2017-11-09 \\
        Michael McIntyre & 2009-12-29 & James White & 2014-05-12 & John Doe & 2017-11-10 \\
        Danny Thomas & 2010-02-04 & Michael Myers & 2014-05-24 & Jean Pierre & 2017-12-06 \\
        Jason Jones & 2010-05-01 & Samuel Johnson & 2014-06-25 & Danny Thomas & 2018-03-22 \\
        Stephen Hill & 2010-06-05 & Jerry Brown & 2014-07-01 & Robert White & 2018-06-11 \\
        Michael White & 2010-06-15 & Justin Johnson & 2014-08-01 & Tafahree Maynard & 2018-10-22 \\
        David Brown & 2010-06-20 & Anthony Brown & 2014-08-23 & Tony Smith & 2018-11-01 \\
        David Smith & 2010-09-17 & Roshad McIntosh & 2014-08-26 & Isaiah Thomas & 2019-02-02 \\
        Robert Thomas & 2010-11-08 & Christopher Anderson & 2014-11-03 & Corey Johnson & 2019-02-04 \\
        Stephen Lee & 2010-12-15 & James Allen & 2015-02-07 & Jason Williams & 2019-03-14 \\
        Eric Williams & 2011-01-12 & Paul Anderson & 2015-04-04 & Daniel Warren & 2019-05-17 \\
        Michael Moore & 2011-02-18 & Walter Scott & 2015-04-04 & Ronald Davis & 2019-09-15 \\
        Larry Brown & 2011-02-27 & Marcus Wheeler & 2015-05-20 & Jordan Griffin & 2019-09-19 \\
        Aaron Williams & 2011-03-17 & James Brown & 2015-08-29 & Lamar Alexander & 2019-12-05 \\
        John White & 2011-04-15 & Jason Day & 2015-10-12 & David Irving & 2020-02-26 \\
        Charles Smith & 2011-04-16 & Tiara Thomas & 2015-11-18 & Robert Johnson & 2020-05-16 \\
        Henry Jones & 2011-05-08 & Robert Covington & 2016-03-03 & William Johnson & 2020-05-21 \\
        William Cooper & 2011-06-18 & James Brown & 2016-03-31 & Michael Thomas & 2020-06-11 \\
        Craig Campbell & 2011-07-06 & Michael Johnson & 2016-06-01 & Paul Williams & 2020-07-07 \\
        Lee Dixon & 2012-01-06 & John Williams & 2016-06-12 & David Brooks & 2020-07-24 \\
        Johnny Wright & 2012-01-10 & Michael Moore & 2016-06-13 & Julian Lewis & 2020-08-07 \\
        Travis Williams & 2012-04-19 & Andre Johnson & 2016-07-09 & Brandon Milburn & 2020-11-11 \\
        Michael Moore & 2012-06-23 & Alvin Ray & 2016-07-25 & James Johnson & 2021-04-07 \\
        Xavier Johnson & 2013-01-04 & Jeffrey Smith & 2016-07-28 & & \\
    \end{tabularx}
    \end{scriptsize}
    \caption{\emph{Names excluded because they received measurable attention in the ten days prior to death.} A person is said to have received measurable attention in the ten days prior to their death if their name was among the top million 2-grams in the ten days prior. This indicates that the name may be shared with another prominent figure, and so they excluded from the analysis. Note that ``Walter Scott" was later added manually because the attention given to the victim of police violence greatly exceeded other usages of the name.}
    \label{tab:anomalies}
\end{table*}

\setlength{\tabcolsep}{12pt}
\begin{table}[!htp]
    \centering
    \begin{scriptsize}
    \begin{tabular}{l|l|l}
        \textbf{Name} & \textbf{Date} & \textbf{Justification} \\
        \hline
        \hline
        Eric Reid & 2009-03-31 & American football player \\
        Darren Wilson & 2009-04-09 & Police officer who killed Michael Brown \\
        Kevin White & 2009-11-15 & American football player \\
        Eddie Jones & 2010-04-28 & Australian rugby coach \\
        Michael Smith & 2012-01-01 & American sports journalist \\
        Bobby Moore & 2012-08-12 & English soccer player \\
        Justin Turner & 2012-11-02 & American baseball player \\
        Eddie Jones & 2012-12-23 & Australian rugby coach \\
        James Anderson & 2013-01-27 & English cricketer \\
        Jeremy Hill & 2013-05-21 & American football player \\
        James Anderson & 2015-09-25 & English cricketer \\
        Thomas Lane & 2016-02-22 & Police officer involved in case of George Floyd \\
    \end{tabular}
    \end{scriptsize}
    \caption{\emph{Manually excluded names for disambiguation purposes.} The automatic name disambiguation process (see Supplementary Table~\ref{tab:anomalies}) misses some cases where most uses of the name on Twitter clearly relate to someone other than the victim of police violence. We remove these manually.}
    \label{tab:manuals}
\end{table}

\begin{table*}[!htp]
    \centering
    %\begin{scriptsize}
    \begin{tabularx}{\textwidth}{ll|ll|ll}
        \textbf{Date} & \textbf{Location}  & \textbf{Date} & \textbf{Location} & \textbf{Date} & \textbf{Location}  \\
        \hline \hline
        2009-05-27 & Chicago, IL & 2012-11-22 & Detroit, MI & 2018-03-05 & Kansas City, MO \\
        2011-03-17 & Fresno, CA & 2013-05-25 & Baltimore, MD & 2019-01-16 & Troy, AL \\
        2011-03-27 & Chicago, IL & 2013-08-20 & Los Angeles, CA & 2020-07-23 & Detroit, MI \\
        2011-10-08 & Baltimore, MD & 2016-03-15 & Tucson, AZ & 2020-11-15 & Inglewood, CA \\
        2012-03-31 & Chicago, IL & 2017-11-15 & Jackson, MS & &
    \end{tabularx}
    %\end{scriptsize}
    \caption{\emph{Cases where the name of the victim was withheld.} Some names in the Fatal Encounters database are ``withheld by police"; therefore, we cannot determine their usage on Twitter. We list the date and location of these cases.}
    \label{tab:unknowns}
\end{table*}

\clearpage

    We expand the dataset with 15 additional names that are not included in the database and their dates of death to reflect key events relating to police brutality and anti-Black violence (giving a final count of 3,737 names). Because these names are chosen manually, they exhibit a relatively high level of attention. See Table~\ref{tab:manual} for a full list of manually selected names in the database.
    \\

\setlength{\tabcolsep}{12pt}
\begin{table*}[!htp]
        \centering
        \begin{scriptsize}
        \begin{tabular}{l|l|l}
            \textbf{Name} & \textbf{Date} & \textbf{Justification}\\
            \hline\hline
            Emmett Till 
                & 1955-08-28 %August 28, 1955 
                & Death before timeframe \\
            Rodney King 
                & 1991-03-03 %March 3, 1991 
                & Injury before timeframe \\
            Amadou Diallo 
                & 1999-02-04 % February 4, 1999 
                & Death before timeframe \\
            Sean Bell 
                & 2006-11-25 %November 25, 2006 
                & Death before timeframe \\
            Trayvon Martin
                & 2012-02-26 %February 26, 2012 
                & Non-police-involved death \\
            Jordan Davis 
                & 2012-11-23 %November 23, 2012 
                & Non-police-involved death \\
            Natasha McKenna 
                & 2015-02-08 %February 8, 2015 
                & Death while in police custody \\
            Walter Scott 
                & 2015-05-04 %April 4, 2015 
                & Measurable attention to name prior to death \\
            Kalief Browder 
                & 2015-06-06 %June 6, 2015 
                & Suicide related to racism in criminal justice system \\
            Sandra Bland 
                & 2015-07-13 %July 13, 2015 
                & Suicide while in police custody \\
            Terrence Crutcher 
                & 2016-09-16 %September 16, 2016 
                & Commonly misspelled name \\
            Ahmaud Arbery 
                & 2020-02-23 %February 23, 2020 
                & Non-police-involved death \\
            Oluwatoyin Salau 
                & 2020-06-06 %June 6, 2020 
                & Non-police-involved death \\
            Robert Fuller 
                & 2020-06-10 %June 10, 2020 
                & Ruled suicide under suspicious circumstances \\
            Jacob Blake 
                & 2020-08-23 %August 23, 2020 
                & Notable injury
        \end{tabular}
        \end{scriptsize}
        \caption{
            \emph{Names that were manually added to the analysis.}
        }
        \label{tab:manual}
    \end{table*}
\setlength{\tabcolsep}{2pt}

    \clearpage

\section{Word Contributions to Expressed Happiness}

    In the main text, we find that the happiness expressed on Twitter following George Floyd's death reached historically low levels. This is based off a dictionary-based sentiment analysis approach. We use the labMT sentiment dictionary~\cite{dodds2011temporal}, which assigns scores to words based on how much ``happiness'' is associated with them. We use an updated version of the dictionary that also includes scores for words related to the pandemic and COVID-19. To calculate the average expressed happiness, we first count how often all the words in the labMT dictionary are used in tweets for a day or period of interest. These counts are treated as a single, large bag of words. We then use the labMT scores to calculate the weighted average, which is the average happiness. 
    
    The average happiness is just one number. We can better understand why it dropped in the wake of George Floyd's death by looking at how particular words contributed to that drop. To do so, we distinguish between different ways that a word can contribute. A word can contribute to the decrease in happiness if it is a \emph{relatively negative} word ($-$) that is used \emph{more frequently} ($\uparrow$). It can also contribute to the decrease if it is a \emph{relatively positive} word ($+$) that is used \emph{less} ($\downarrow$). On the other hand, a word can be counter to the decrease in happiness if it is a \emph{relatively positive} word ($+$) that was used more ($\uparrow$), or a \emph{relatively negative} word ($-$) that was used less ($\downarrow$). We treat the week prior to George Floyd's death as a reference period. We say that a word is \emph{relatively} positive or negative if its labMT score is higher or lower than the average happiness of the reference period. We visualize the word contributions as word shift graphs \cite{dodds2011temporal,gallagher2021generalized}, vertical bar charts showing the magnitude of how much each word contributes to the decrease in happiness following George Floyd's death. Deep yellow bars indicate relatively positive words that were used more ($+\uparrow$), deep blue bars indicate negative words that were used more ($-\uparrow$), light yellow bars indicate positive words that were used less ($+\downarrow)$, and light blue bars indicate negative words that were used less $(-\downarrow)$. The relative magnitude of each type of word contribution is shown at the top of the word shift graphs.

\begin{figure*}[!htp]
          \centering
          \includegraphics[scale=0.58]{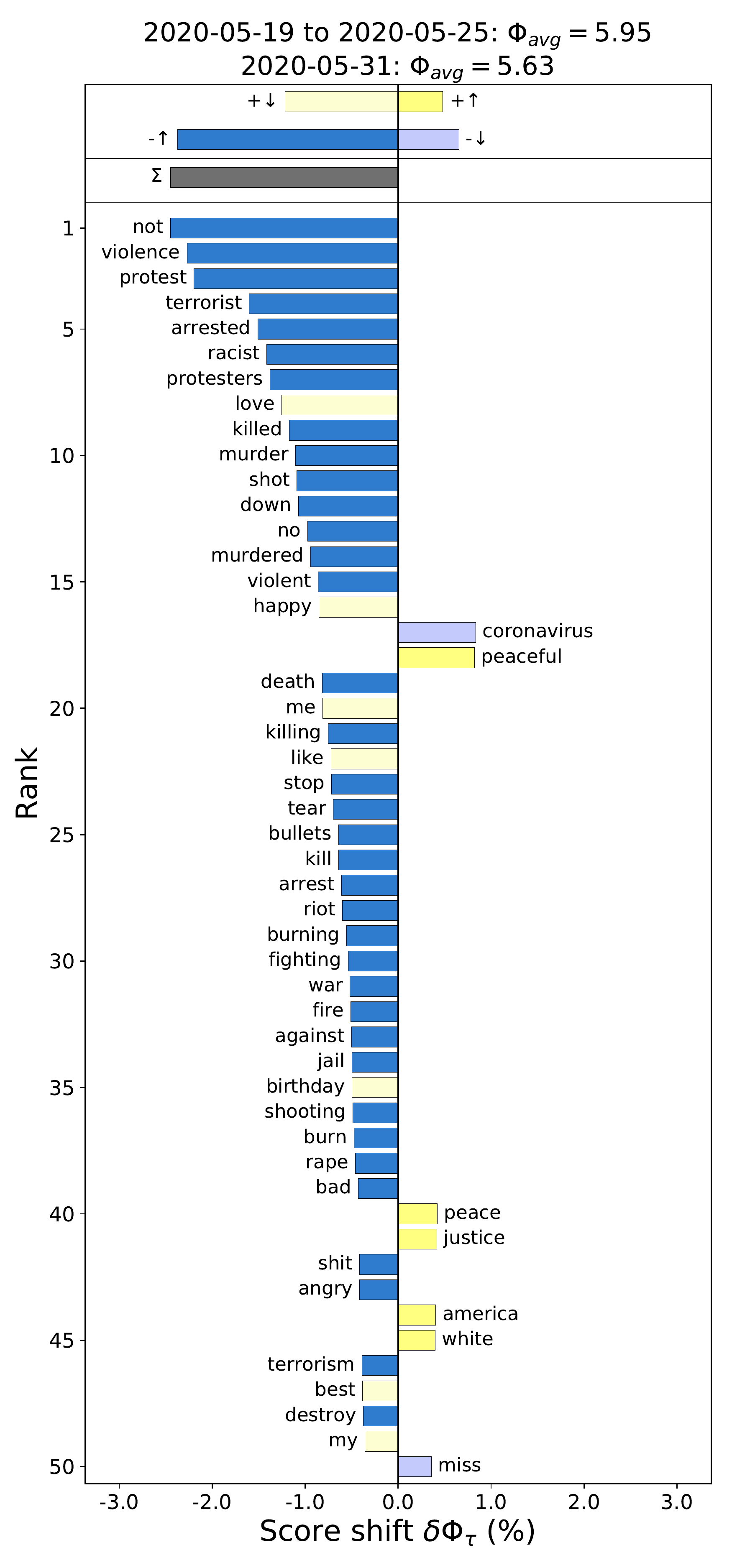}
          \caption{
            \emph{Words that contributed to the ``saddest'' day recorded in Twitter's history, May 31st, 2020.} The week prior to George Floyd's death is used as a reference period.
          }
          \label{fig:saddestday}
    \end{figure*}

    Figure~\ref{fig:saddestday} shows the word shift graph for May 31, 2020, the ``saddest'' day (day with the least happiness expressed) ever recorded by the Hedonometer, about a week after George Floyd's death. By the top of Fig.~\ref{fig:saddestday}, we can see that increases in the frequency of negative words form the largest contribution to the change in happiness on this day. Increases in positive words and decreases in negative words make up the smallest contributions, and they are about equal. We observe that words such as ``terrorist," ``protest," ``violence," and ``racist" make large contributions as negative words that were used more. They relate to the protests following George Floyd's death and general conversation surrounding the topic of racism. Relatively happy words relating to ``peace" contribute to shifting the happiness change in the other direction. Many coronavirus-related terms, which are relatively negative, were used less frequently, having been displaced by the protest- and racism-related terms.

\begin{figure*}[!htp]
          \centering
          \includegraphics[scale=0.58]{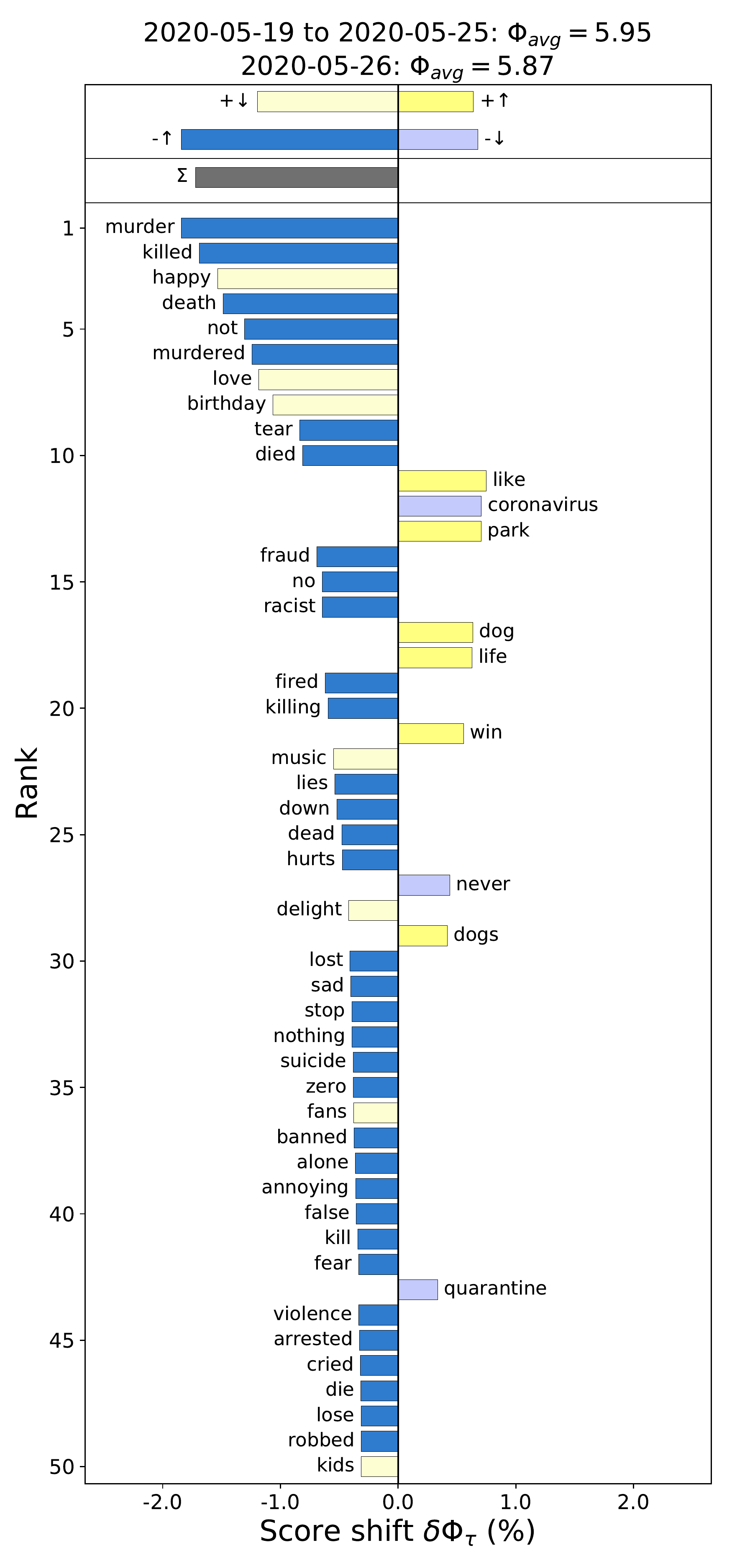}
          \caption{
            \emph{Words that contributed to decreased happiness on May 26th, 2020, the day following George Floyd's death.} The week prior to George Floyd's death is used as a reference period.
          }
          \label{fig:wordshiftmay26}
    \end{figure*}

\begin{figure*}[!htp]
          \centering
          \includegraphics[scale=0.58]{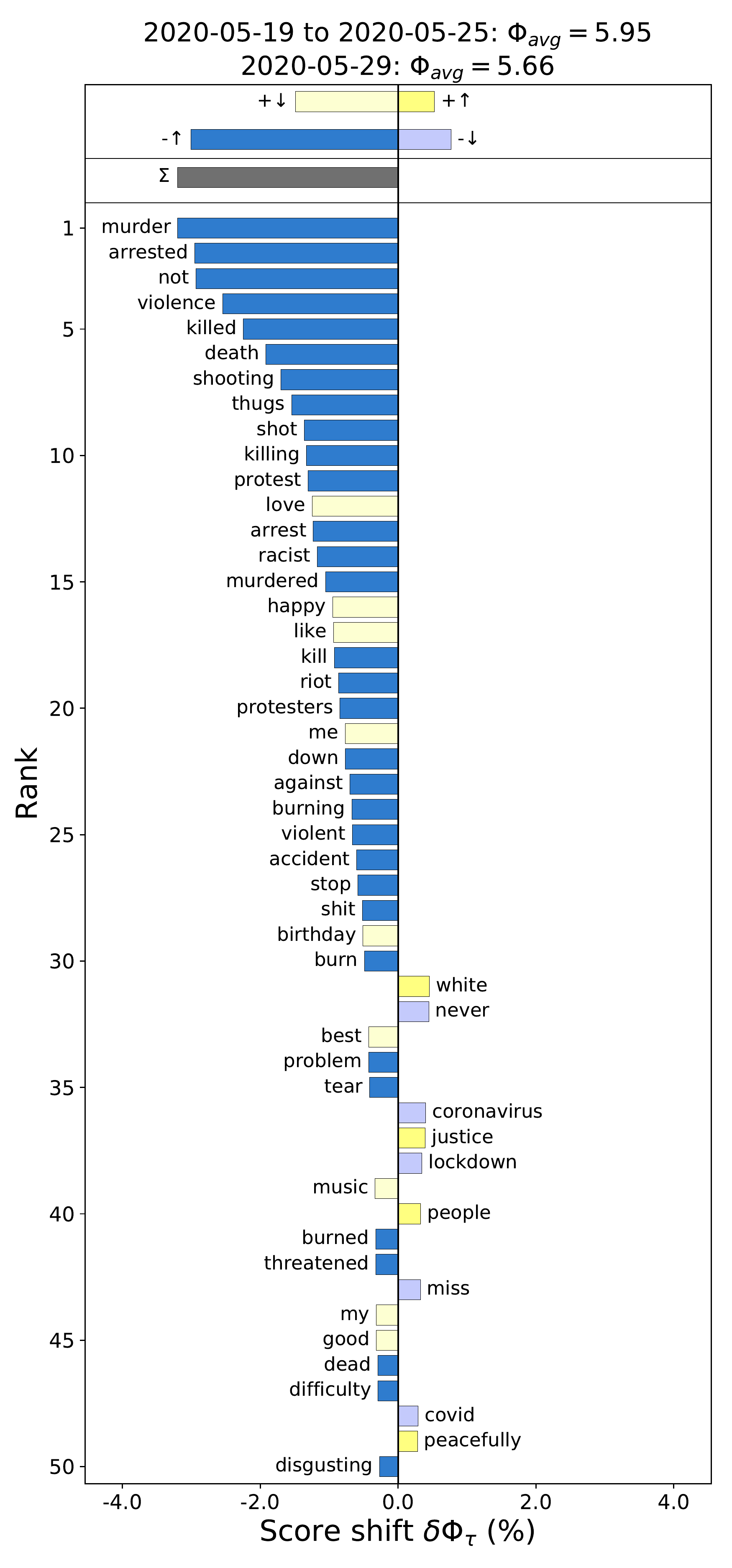}
          \caption{
            \emph{Words that contributed to decreased happiness on May 29th, 2020.} The week prior to George Floyd's death is used as a reference period.
          }
          \label{fig:wordshiftmay29}
    \end{figure*}

\begin{figure*}[!htp]
          \centering
          \includegraphics[scale=0.58]{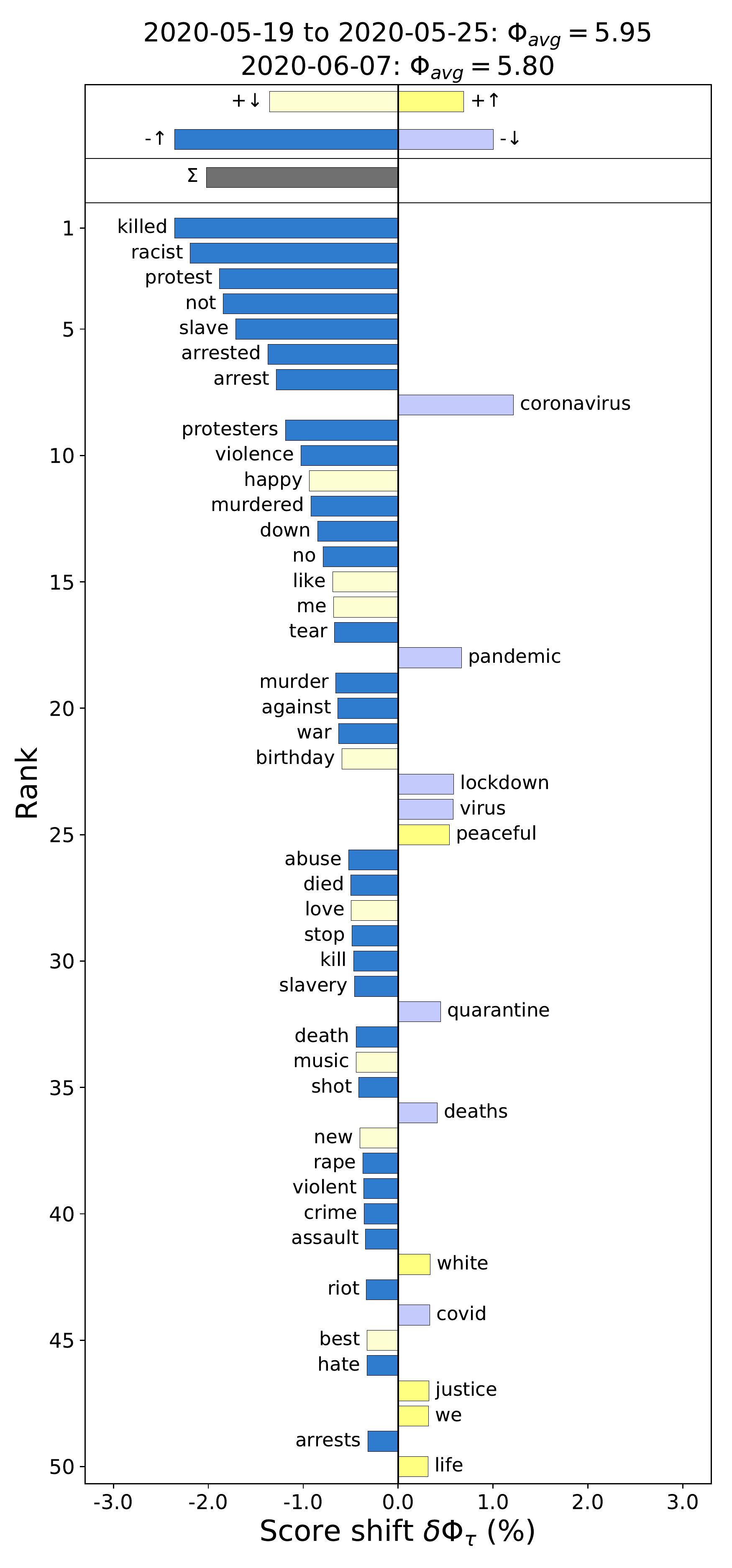}
          \caption{
            \emph{Words that contributed to decreased happiness on June 7th, 2020, the end of the spike period.} The week prior to George Floyd's death is used as a reference period.
          }
          \label{fig:wordshiftjune7}
    \end{figure*}

    We also show word shift graphs for three other specific days: the day following George Floyd's death (May 26th, 2020, see Figure~\ref{fig:wordshiftmay26}), several days following his death (May 29th, see Figure~\ref{fig:wordshiftmay29}), and the final day of the spike period (June 7th, see Figure~\ref{fig:wordshiftjune7}). Immediately following his death, we see increased use of negative words relating to that death, like ``murder,'' ``killed,'' and ``died'' as well as terms related to the specific incident like ``fraud'' and ``down.'' Note the words ``dog,'' ``dogs,'' and ``park'': on the day of George Floyd's death, May 25th, there was also a video-recorded incident in which a white woman walking a dog called the police on a Black man who was simply birdwatching in Central Park in New York City. By May 29th, those negative words have shifted to more protest-related terms, like ``protesters,'' ``riot,'' ``thug,'' and ``tear.'' Note, on both May 26th and the 29th, panndemic related words like ``coronavirus,'' ``lockdown,'' and ``quarantine'' were used less. Also by the 29th, we see some words being used more, like ``justice'' and ``white,'' suggesting shifts to broader discussions of racial justice. These words still see increase usage by the end of the spike period, relative to the week prior to the spike, and other words like ``slave'' and ``slavery'' were used more.

\clearpage    

\section{Attention Following the Death of George Floyd}     

\begin{table*}[!htp]
        \centering
        \begin{tabular}{c|c|c|c|c}
            \textbf{Spike Period}
            & \shortstack{Number of Names with\\Increased Attention}
            & \shortstack{Perc. of Names with No \\Attention $n$ Days Before}
            & \shortstack{Avg. Diff. in Rel. Freq.\\Spike - Before}
            %& $p$-value
            & \shortstack{Avg. Diff. in Rel. Freq.\\After - Before} \\
            %& $p$-value \\%& $\mu_3 - \mu_2$ \\
            \hline
            \hline
            \multicolumn{5}{c}{\textbf{7 days}} \\
            \hline
            Nov. 24--Dec. 8, 2014
                & 87
                & 78.1\%
                & 3.73e-08 
                %& 0.106
                & 8.76e-09 \\
                %& 0.162 \\
            \hline
            Jul. 13--Jul. 26, 2015
                & 46
                & 73.9\%
                & 2.39e-09 
                %& 0.058
                & \textbf{1.36e-09**} \\
                %& 0.010 \\
            \hline
            Jul. 5--Jul. 13, 2016
                & 70
                & 81.4\%
                & \textbf{7.94e-09*}
                %& 0.036
                & 4.55e-09 \\
                %& 0.076 \\
            \hline
            Aug. 12--Aug. 22, 2017
                & 38
                & 89.4\%
                & 8.53e-10
                %& 0.722
                & 4.60e-10\\
                %& 0.863 \\
            \hline
            May 25--Jun. 6, 2020
                & 191
                & 92.6\%
                & \textbf{5.71e-08**}
                %& 0.002
                & 5.21e-08\\
                %& 0.117 \\
            \hline
            \hline
            \multicolumn{5}{c}{\textbf{60 days}} \\
            \hline
            Nov. 24--Dec. 8, 2014
                & 78
                & 56.4\%
                & 3.77e-08
                %& 0.109
                & 2.41e-10\\
                %& 0.898 \\
            \hline
            Jul. 13--Jul. 26, 2015
                & 36
                & 55.5\%
                & -3.62e-11
                %& 0.979
                & -6.64e-10\\
                %& 0.455 \\
            \hline
            Jul. 5--Jul. 13, 2016
                & 66
                & 56.0\%
                & 6.63e-09
                %& 0.075
                & \textbf{1.82e-09**}\\
                %& 0.025 \\
            \hline
            Aug. 12--Aug. 22, 2017
                & 34
                & 32.3\%
                & -3.64e-09
                %& 0.294
                & -4.16e-09\\
                %& 0.105 \\
            \hline
            May 25--Jun. 6, 2020
                & 183
                & 77.0\%
                & \textbf{5.65e-08**}
                %& 0.002
                & 2.45e-08\\
                %& 0.106 \\
            \hline
            \hline
            \multicolumn{5}{c}{\textbf{90 days}} \\
            \hline
            Nov. 24--Dec. 8, 2014
                & 76
                & 47.3\%
                & 3.73e-08
                %& 0.110
                & -6.85e-10\\
                %& 0.695 \\
            \hline
            Jul. 13--Jul. 26, 2015
                & 36
                & 16.6\%
                & -2.39e-09
                %& 0.332
                & -3.34e-09\\
                %& 0.118 \\
            \hline
            Jul. 5--Jul. 13, 2016
                & 62
                & 48.3\%
                & 6.72e-09
                %& 0.067
                & 1.47e-09\\
               % & 0.057 \\
            \hline
            Aug. 12--Aug. 22, 2017
                & 34
                & 27.7\%
                & -2.60e-09
                %& 0.322
                & -3.32e-09\\
                %& 0.074 \\
            \hline
            May 25--Jun. 6, 2020
                & 184
                & 72.2\%
                & \textbf{5.81e-08**}
                %& 0.002
                & 1.89e-08
               % & 0.101 \\
        \end{tabular}
        \caption{\emph{Robustness of resurgent attention to past victims of police violence following George Floyd's death}. We consider four periods of spikes in attention relevant to \#BlackLivesMatter: November 24th--December 8th, 2014 (deaths and non-indictments in the cases of Michael Brown, Tamir Rice, and Eric Garner), July 13th--July 26th, 2015 (death of Sandra Bland), July 5th--July 13th, 2016 (deaths of Philando Castile and Alton Sterling), August 12th--August 22nd, 2017 (``Unite the Right'' Charlottesville rally), and May 25th--June 6th, 2020 (death of George Floyd). We vary the period before and after the spike across $n=7, 60$, and $90$ days. The number of names that received increased attention during a spike period is reported, as well as the percentage of those that had not received any measurable attention in the $n$ days prior to the spike. The average change in average relative frequency is calculated for the difference between $n$ days before the spike period and during it, and $n$ days before and after it. Statistical significance is indicated by * for $\alpha=0.05$ and ** for $\alpha=0.01$.}
        \label{tab:resurgence90}
    \end{table*}

    % \begin{table*}[!htp]
    %     \centering
    %     \begin{tabular}{c|c|c|c|c|c|c}
    %         Spike Period & 7 days & $p$-value & 60 days & $p$-value & 90 days & $p$-value \\%& $\mu_3 - \mu_2$ \\
    %         \hline
    %         \hline
    %         Nov. 24--Dec. 8, 2014
    %             & -1.00e-06
    %             & 0.053
    %             & 2.95e-08
    %             & 0.505
    %             & 1.15e-08
    %             & 0.594 \\
    %         \hline
    %         Jul. 13--Jul. 26, 2015
    %             & NaN
    %             & NaN
    %             & -2.83e-08
    %             & 0.564
    %             & 1.22e-08
    %             & 0.452 \\
    %         \hline
    %         Jul. 5--Jul. 13, 2016
    %             & 2.76e-06
    %             & 0.407
    %             & 7.68e-08
    %             & 0.180
    %             & 1.47e-07
    %             & 0.121 \\
    %         \hline
    %         Aug. 12--Aug. 22, 2017
    %             & 0
    %             & NaN
    %             & -1.36e-07
    %             & 0.261
    %             & -8.47e-08
    %             & 0.163 \\
    %         \hline
    %         May 25--Jun. 6, 2020
    %             & 4.63e-06
    %             & 0.482
    %             & 8.70e-07
    %             & 0.298
    %             & 1.21e-06
    %             & 0.237 \\
    %     \end{tabular}
    %     \caption{Differences in means of mean initial frequencies.}
    %     \label{tab:initialdifferencerobust}
    % \end{table*}

        \begin{table*}
        \centering
        \begin{scriptsize}
        \begin{tabularx}{\textwidth}{rC|rC|rC|rC}
            \textbf{Name} & \textbf{Date} & \textbf{Name} & \textbf{Date} & \textbf{Name} & \textbf{Date} & \textbf{Name} & \textbf{Date} \\
            \hline
            \hline
            Sean Bell           & 2009-01-01 & Joshua Johnson    & 2013-05-20 & Sandra Bland        & 2015-07-13 & John Young            & 2018-12-01 \\
            Oscar Grant         & 2009-01-01 & Larry Jackson     & 2013-07-26 & Darrius Stewart     & 2015-07-17 & Christopher Mitchell  & 2018-12-09 \\
            Amadou Diallo       & 2009-01-01 & John Allen        & 2013-08-29 & Samuel DuBose       & 2015-07-19 & Jameek Lowery         & 2019-01-05 \\
            Rodney King         & 2009-01-01 & Brian Nichols     & 2013-09-03 & Christian Taylor    & 2015-08-07 & Jacob Harris          & 2019-01-11 \\
            Emmett Till         & 2009-01-01 & William Brown     & 2013-09-10 & Redel Jones         & 2015-08-12 & Jimmy Atchison        & 2019-01-22 \\
            William Smith       & 2009-01-12 & Jonathan Ferrell  & 2013-09-14 & Jean Charles        & 2015-08-14 & Joshua Williams       & 2019-02-08 \\
            Eric Reid           & 2009-03-31 & Miriam Carey      & 2013-10-03 & Ricky Ball          & 2015-10-16 & Willie McCoy          & 2019-02-09 \\
            Tony Anderson       & 2009-06-24 & William Harvey    & 2013-10-27 & Jamar Clark         & 2015-11-15 & Mario Clark           & 2019-02-14 \\
            Charles Brown       & 2009-07-08 & Robert Brown      & 2013-11-25 & Mario Woods         & 2015-12-02 & Osaze Osagie          & 2019-03-20 \\
            James Miller        & 2009-10-21 & Jason Lewis       & 2013-11-27 & Michael Noel        & 2015-12-21 & Isaiah Lewis          & 2019-04-29 \\
            Kevin White         & 2009-11-15 & Kenneth Herring   & 2013-12-12 & Bettie Jones        & 2015-12-26 & Pamela Turner         & 2019-05-13 \\
            Christopher Wright  & 2010-02-28 & Gregory Hill      & 2014-01-14 & Che Taylor          & 2016-02-21 & Dominique Clayton     & 2019-05-19 \\
            Aiyana Jones        & 2010-05-16 & Jordan Baker      & 2014-01-16 & Matthew Tucker      & 2016-05-04 & Miles Hall            & 2019-06-02 \\
            Kemp Yarborough     & 2011-03-08 & James Norris      & 2014-02-05 & Henry Green         & 2016-06-06 & Ryan Twyman           & 2019-06-06 \\
            Jerry Moore         & 2011-06-17 & Yvette Smith      & 2014-02-16 & Alton Sterling      & 2016-07-05 & JaQuavion Slaton      & 2019-06-09 \\
            Maurice Hampton     & 2011-06-30 & Victor White      & 2014-03-03 & Philando Castile    & 2016-07-06 & Brandon Webber        & 2019-06-12 \\
            Kenneth Chamberlain & 2011-11-19 & Dontre Hamilton   & 2014-04-30 & Micah Johnson       & 2016-07-07 & Eric Logan            & 2019-06-16 \\
            Malik Williams      & 2011-12-10 & Pearlie Golden    & 2014-05-06 & Korryn Gaines       & 2016-08-01 & Isak Aden             & 2019-07-02 \\
            Wayne Williams      & 2011-12-22 & Eric Harris       & 2014-06-15 & Jamarion Robinson   & 2016-08-05 & Sean Rambert          & 2019-07-09 \\
            Lawrence Jones      & 2011-12-23 & Lavon King        & 2014-06-24 & Donta Taylor        & 2016-08-25 & Elijah McClain        & 2019-08-24 \\
            Stephon Watts       & 2012-02-01 & Eric Garner       & 2014-07-17 & Terrence Sterling   & 2016-09-11 & Byron Williams        & 2019-09-05 \\
            Ramarley Graham     & 2012-02-02 & John Crawford     & 2014-08-05 & Tyre King           & 2016-09-14 & Bennie Branch         & 2019-09-08 \\
            Trayvon Martin      & 2012-02-26 & Michael Brown     & 2014-08-09 & Terence Crutcher    & 2016-09-16 & Atatiana Jefferson    & 2019-10-12 \\
            Wendell Allen       & 2012-03-07 & Ezell Ford        & 2014-08-11 & Terrence Crutcher   & 2016-09-16 & Christopher Whitfield & 2019-10-14 \\
            Shereese Francis    & 2012-03-15 & Dante Parker      & 2014-08-12 & Keith Scott         & 2016-09-20 & Dana Fletcher         & 2019-10-27 \\
            Rekia Boyd          & 2012-03-21 & Michelle Cusseaux & 2014-08-14 & Deborah Danner      & 2016-10-18 & David Smith           & 2019-10-28 \\
            Kendrec McDade      & 2012-03-24 & Kajieme Powell    & 2014-08-19 & Quanice Hayes       & 2017-02-09 & Mark Sheppard         & 2019-11-15 \\
            James Weldon        & 2012-04-10 & Darrien Hunt      & 2014-09-10 & Cordale Handy       & 2017-03-15 & Ariane McCree         & 2019-11-23 \\
            Alan Blueford       & 2012-05-06 & Laquan McDonald   & 2014-10-20 & Jordan Edwards      & 2017-04-29 & Michael Dean          & 2019-12-02 \\
            Derrick Gaines      & 2012-06-05 & Aura Rosser       & 2014-11-09 & Terrell Johnson     & 2017-05-10 & Cameron Lamb          & 2019-12-03 \\
            Christopher Brown   & 2012-06-13 & Tanisha Anderson  & 2014-11-13 & Andrew Kearse       & 2017-05-11 & Jamee Johnson         & 2019-12-14 \\
            Shantel Davis       & 2012-06-14 & Akai Gurley       & 2014-11-20 & Marc Davis          & 2017-06-02 & Darius Tarver         & 2020-01-21 \\
            Matthew Henderson   & 2012-07-13 & Tamir Rice        & 2014-11-22 & Charleena Lyles     & 2017-06-18 & William Green         & 2020-01-27 \\
            Alesia Thomas       & 2012-07-22 & Rumain Brisbon    & 2014-12-02 & Corey Mobley        & 2018-01-23 & Manuel Ellis          & 2020-03-03 \\
            Trevor Taylor       & 2012-07-31 & Jerame Reid       & 2014-12-30 & Ronell Foster       & 2018-02-13 & Donnie Sanders        & 2020-03-12 \\
            Bobby Moore         & 2012-08-12 & Natasha McKenna   & 2015-02-08 & Stephon Clark       & 2018-03-18 & Breonna Taylor        & 2020-03-13 \\
            Anthony Anderson    & 2012-09-21 & Anthony Hill      & 2015-03-09 & Shukri Said         & 2018-04-28 & Joshua Ruffin         & 2020-04-08 \\
            Jordan Davis        & 2012-11-23 & Nicholas Thomas   & 2015-03-24 & Marcus-David Peters & 2018-05-14 & Desmond Franklin      & 2020-04-09 \\
            Malissa Williams    & 2012-11-29 & Mya Hall          & 2015-03-30 & Marqueese Alston    & 2018-06-12 & Steven Taylor         & 2020-04-18 \\
            Shelly Frey         & 2012-12-06 & Phillip White     & 2015-03-31 & Antwon Rose         & 2018-06-19 & Jonas Joseph          & 2020-04-28 \\
            James Anderson      & 2013-01-27 & Walter Scott      & 2015-04-04 & Daniel Hambrick     & 2018-07-26 & Denzel Taylor         & 2020-04-29 \\
            George Walker       & 2013-02-01 & Don Smith         & 2015-04-09 & Diamond Ross        & 2018-08-18 & Said Joquin           & 2020-05-01 \\
            Kayla Moore         & 2013-02-12 & Freddie Gray      & 2015-04-12 & Botham Jean         & 2018-09-06 & Finan Berhe           & 2020-05-07 \\
            Christopher Taylor  & 2013-02-21 & David Felix       & 2015-04-25 & Willie Simmons      & 2018-09-28 & Yassin Mohamed        & 2020-05-09 \\
            Kimani Gray         & 2013-03-09 & Alexia Christian  & 2015-04-30 & Patrick Kimmons     & 2018-09-30 & Maurice Gordon        & 2020-05-23 \\
            Kenneth Williams    & 2013-05-01 & Brendon Glenn     & 2015-05-05 & Jemel Roberson      & 2018-11-11 &                       &            \\
            Terrance Franklin   & 2013-05-10 & Kalief Browder    & 2015-06-06 & Emantic Bradford    & 2018-11-22 &                       &            
        \end{tabularx}
        \end{scriptsize}
        \caption{\emph{Names of those who received increased attention during the spike following George Floyd's death.} A name received increased attention if its the mean relative frequency from May 25 to June 7, 2020 was greater than its mean relative frequency from April 25 to May 24, 2020.}
        \label{tab:namelist1}
    \end{table*}

\clearpage

\section{Long-Term Attention and Amplification}

    \begin{figure*}[h!tp]
      \centering	
        \includegraphics[width=\textwidth]{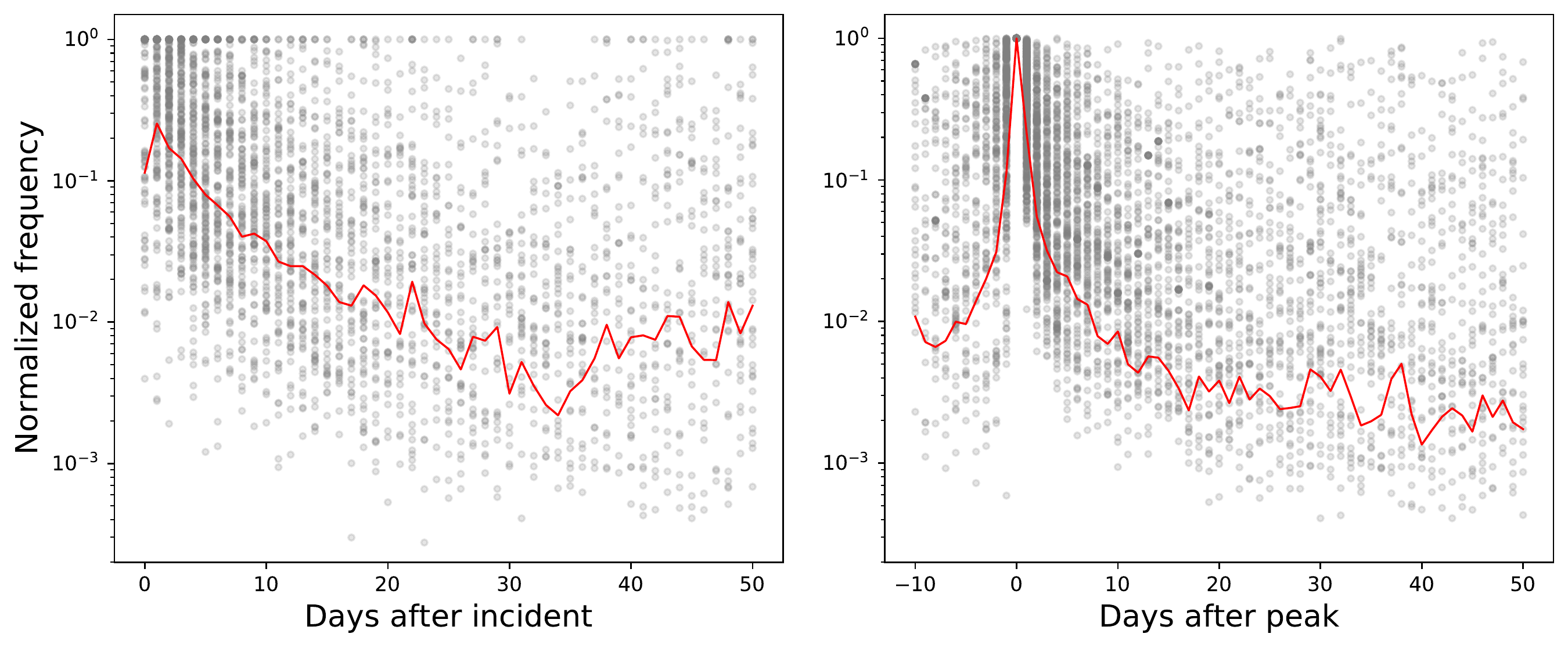}  
      \caption{
        \emph{Attention decay relative to the death of a victim and the day that their name received peak attention.} Scatter plots of normalized attention $\widehat{\ngramprop}$ for all Black victims in the combined database anchored to the first 50 days after the date of death (left) and peak rank (right). 
        We exclude names that did not receive measurable attention during the specified period, i.e. did not appear in the top million 2-grams. 
        The red line plot shows the mean of the included values on each day.
        Because of the log-scaled y-axis, points with a normalized frequency of 0 are not visible but are included in the mean. 
        %See Appendix Fig.~\ref{fig:attentiondecaygender} for the same figure separated by gender.
      }
      \label{fig:attentiondecay}
    \end{figure*}

    \begin{figure*}[h!tp]
      \centering
        \includegraphics[width=\textwidth]{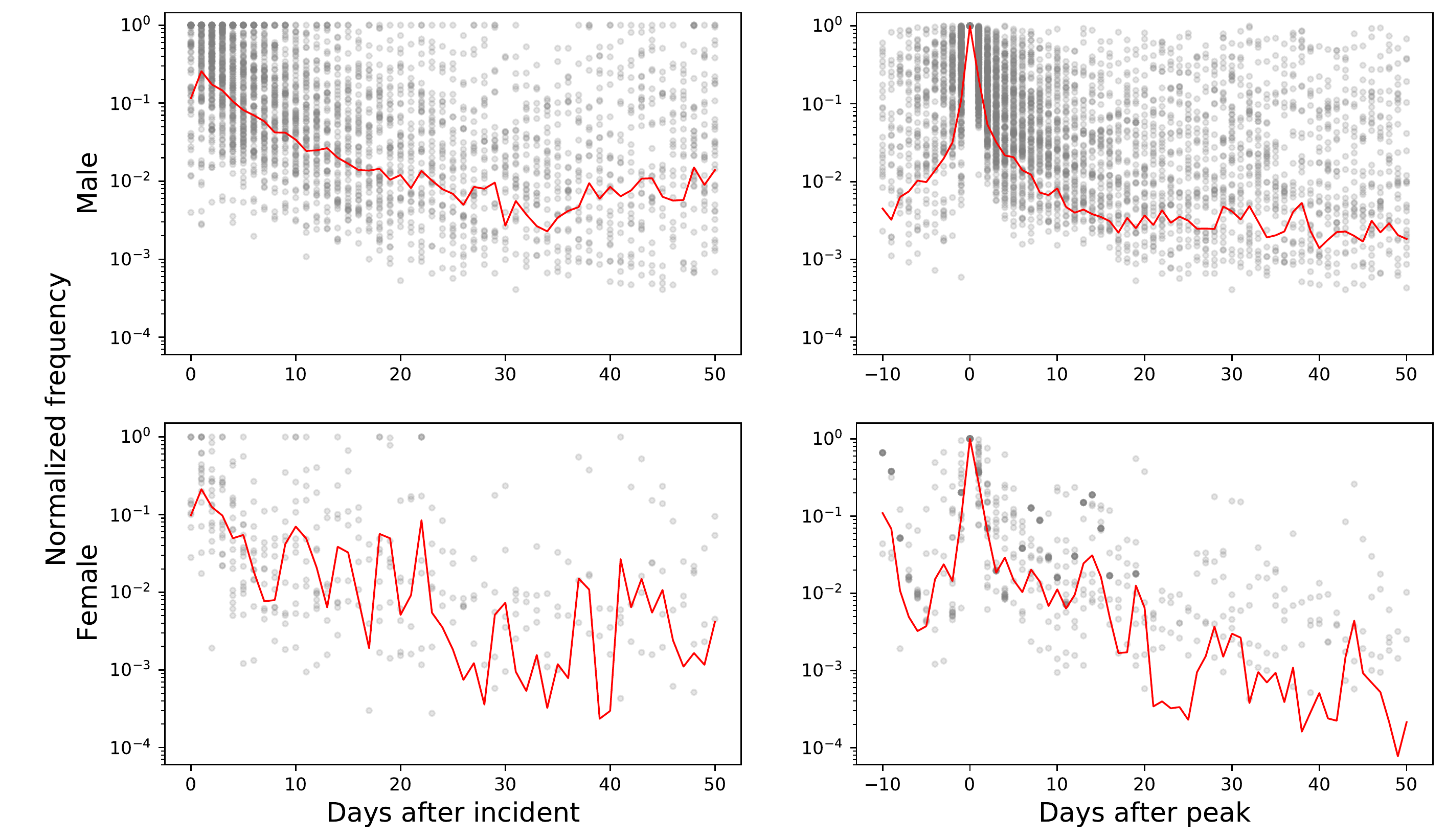}
      \caption{
        \emph{Attention decay by gender.} See Supplementary Figure~\ref{fig:attentiondecay} for details.
      }
      \label{fig:attentiondecaygender}
    \end{figure*}

    \begin{figure*}[h!tp]
      \centering
        \includegraphics[scale=0.65]{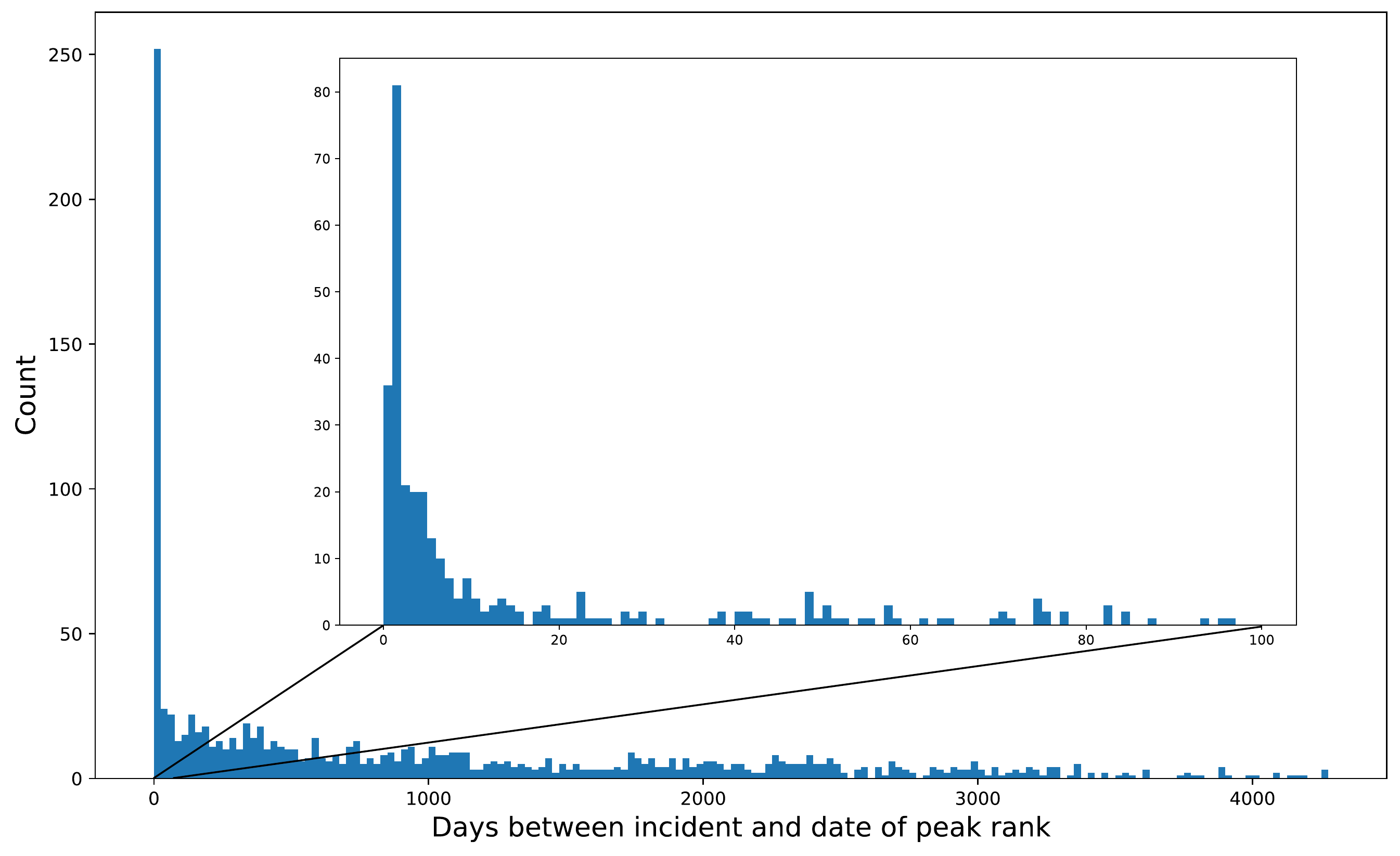}
      \caption{
        \emph{Histogram of delay between date of incident and date of peak rank.} The main figure uses 25-day bins, while the inset uses 1-day bins from 0 to 100 days between incident and peak rank. Short delays are more common than long delays, with 0 and 1 days between the incident and the peak rank of a name being the most common. This provides evidence that most names in our analyses are properly disambiguated.
      }
      \label{fig:peakdelayhist}
    \end{figure*}
    
    \clearpage

    \begin{figure*}[h!tp]
      \centering
        \includegraphics[scale=0.55]{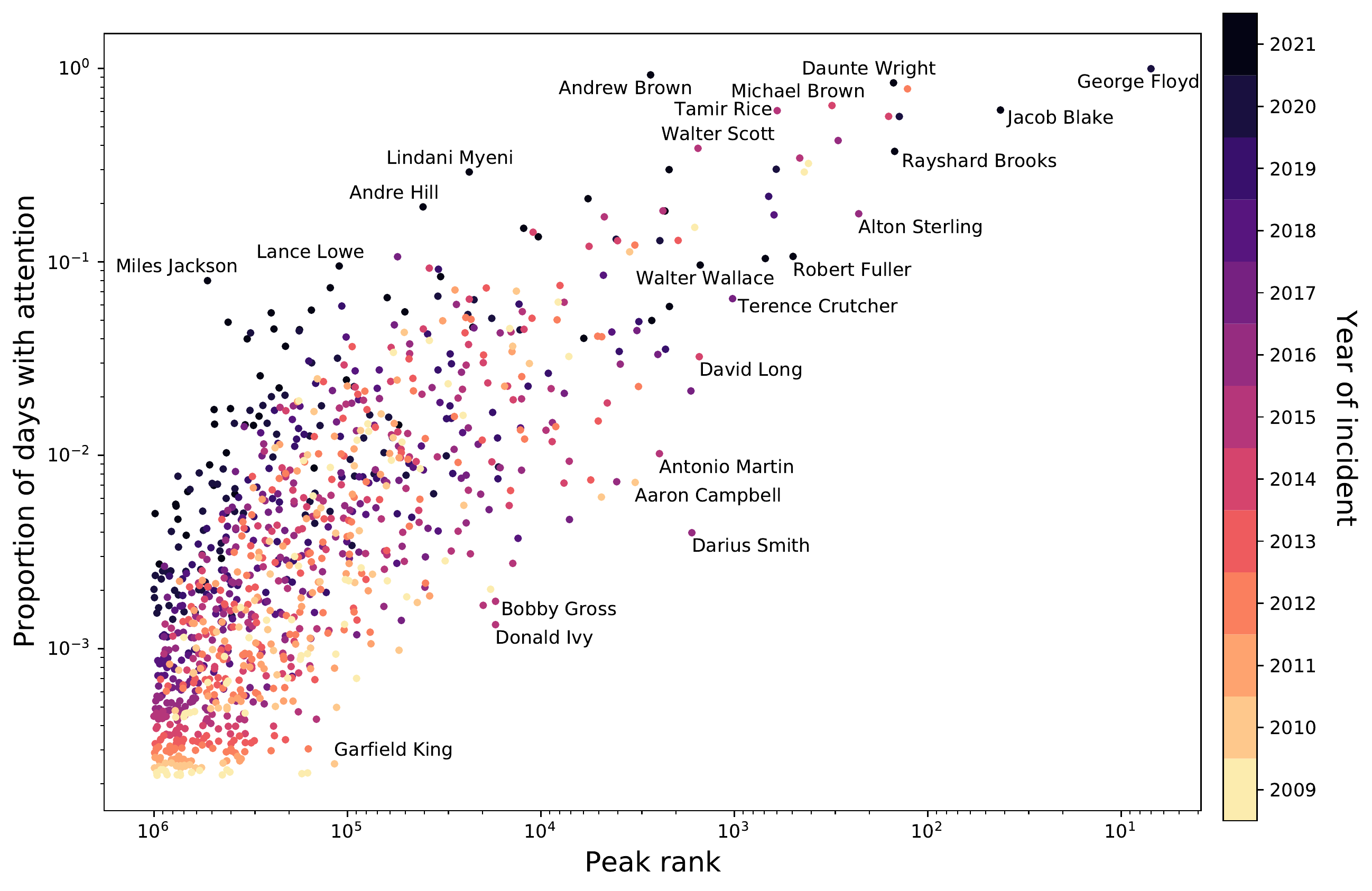}
      \caption{
        \emph{Proportion of days with measurable attention and peak rank of Black male victims of fatal police violence.} See Figure~\ref{fig:namerank} for details.
      }
      \label{fig:namerankmale}
    \end{figure*}
    
    \begin{figure*}[h!tp]
      \centering
        \includegraphics[scale=0.6]{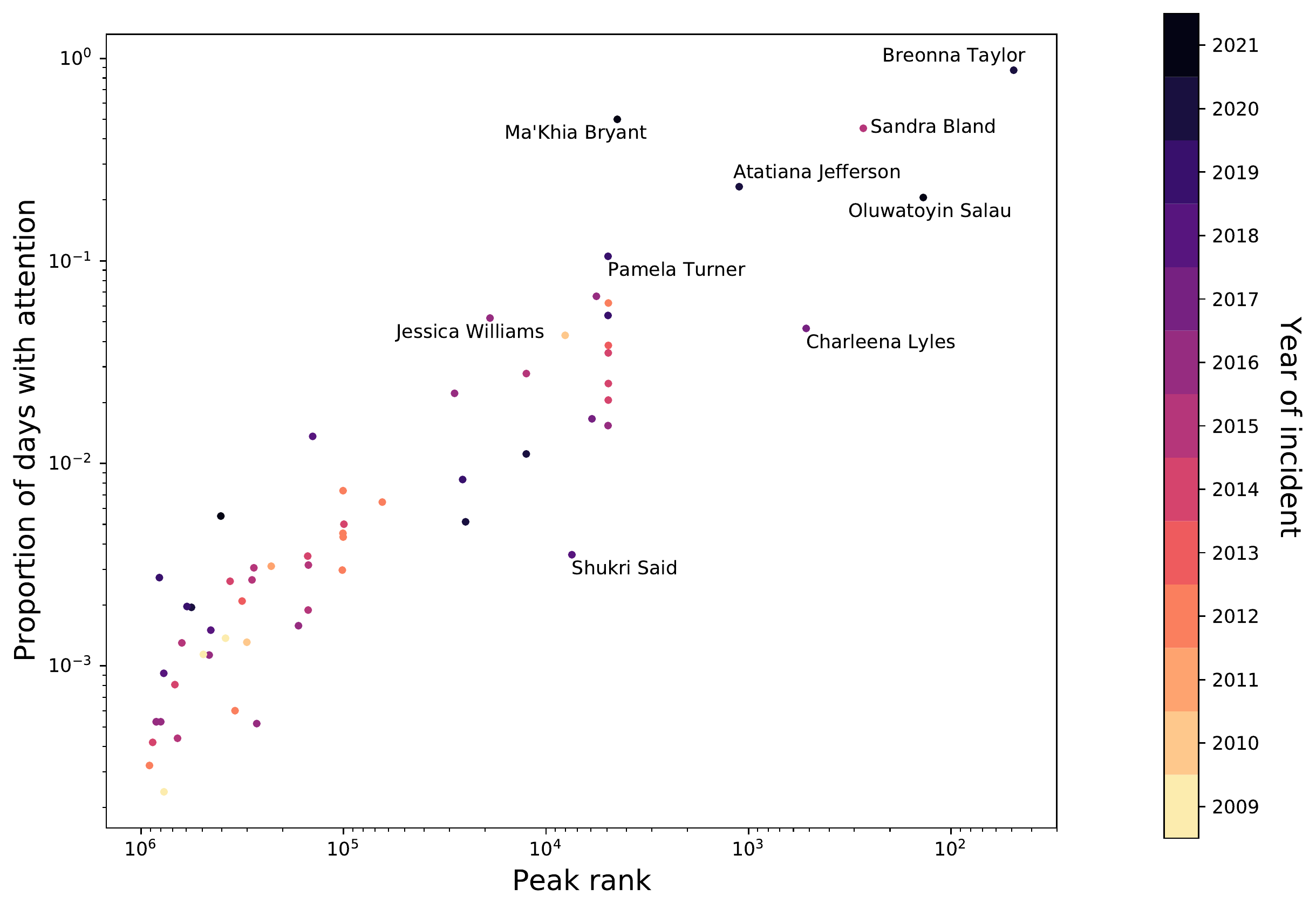}
      \caption{
        \emph{Proportion of days with measurable attention and peak rank of Black female victims of fatal police violence.} See Figure~\ref{fig:namerank} for details.
      }
      \label{fig:namerankfemale}
    \end{figure*}

    \begin{figure*}[h!tp]
      \centering
        \includegraphics[width=\textwidth]{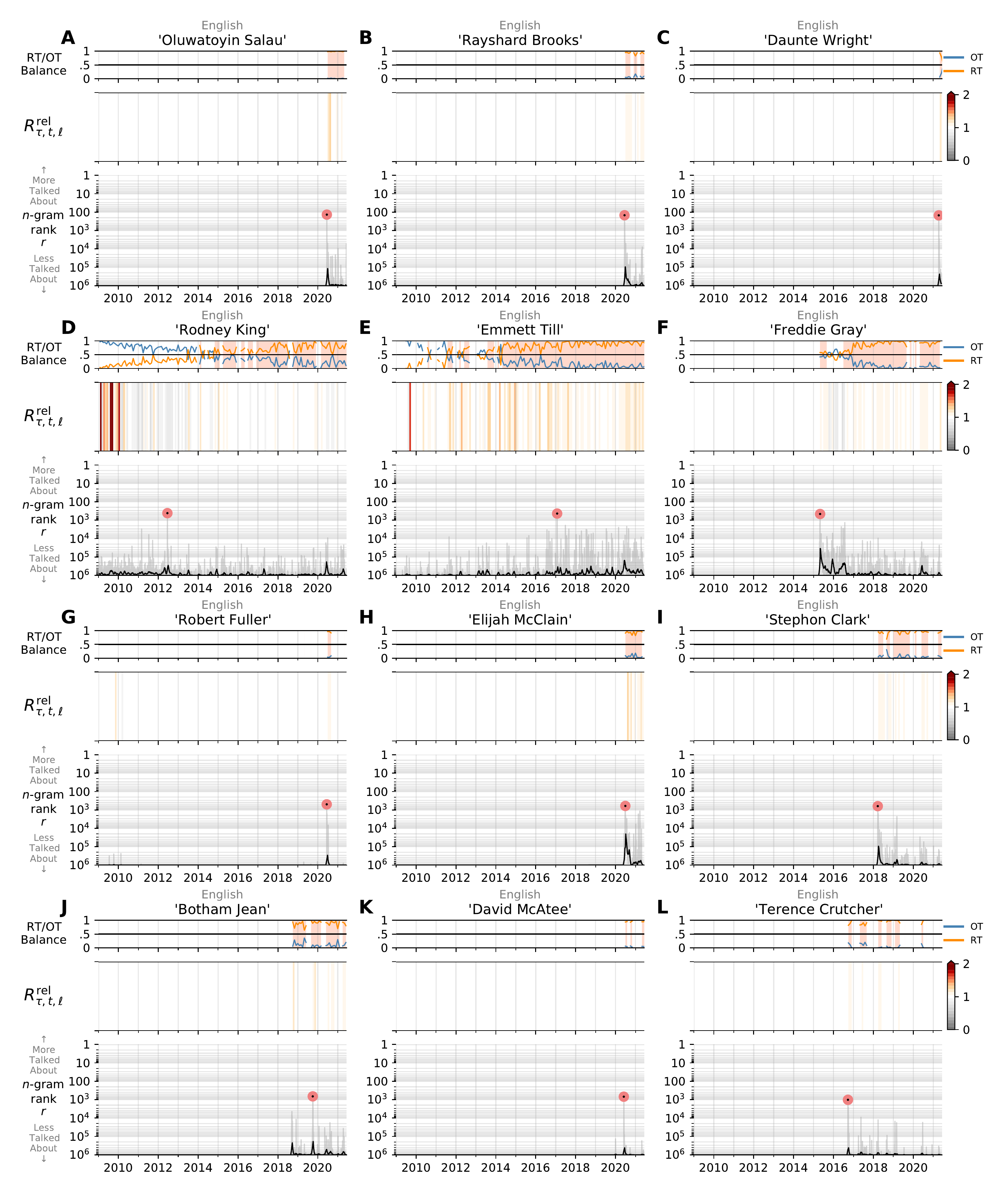}
      \caption{\emph{Amplification of additional emblematic names of Black Lives Matter.} We see patterns similar to those observed in Figure~\ref{fig:namecontagiogramsmain}, such as high relative social amplification ($>1$) and dual peaks of attention.}
      \label{fig:namecontagiogramssupp}
    \end{figure*}

    \begin{figure*}[h!tp]
      \centering
        \includegraphics[scale=0.57]{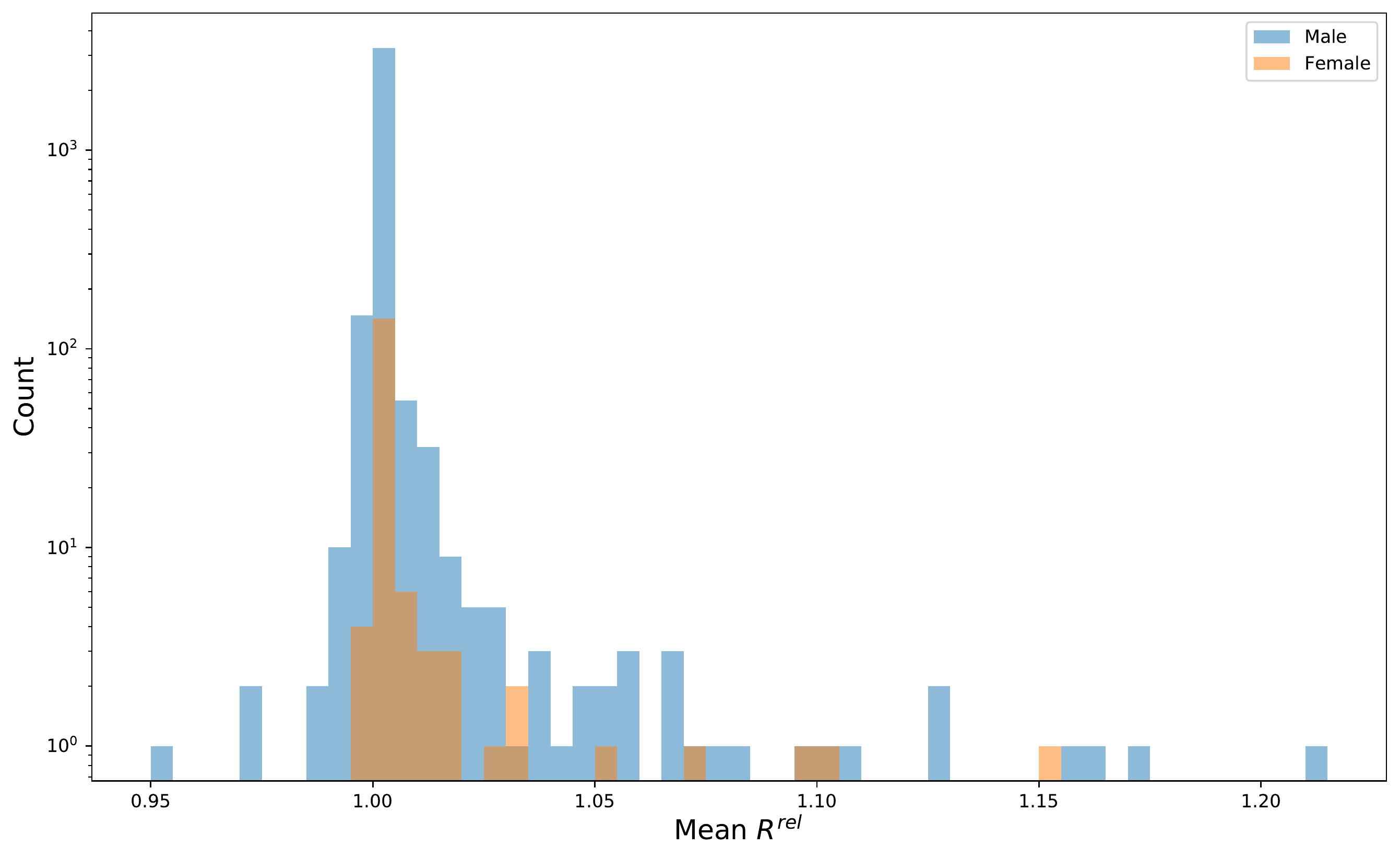}
      \caption{
        \emph{Histogram of mean relative social amplification across names, separated by gender.} The bin size is 0.005. Although there are many more men than women in our analysis, their distributions of mean relative social amplification are similar. Many are concentrated around 1 and the distributions skew right, with more values above 1 than below it.
      }
      \label{fig:meanratiohist}
    \end{figure*}

    \begin{figure*}[h!tp]
      \centering
        \includegraphics[scale=0.57]{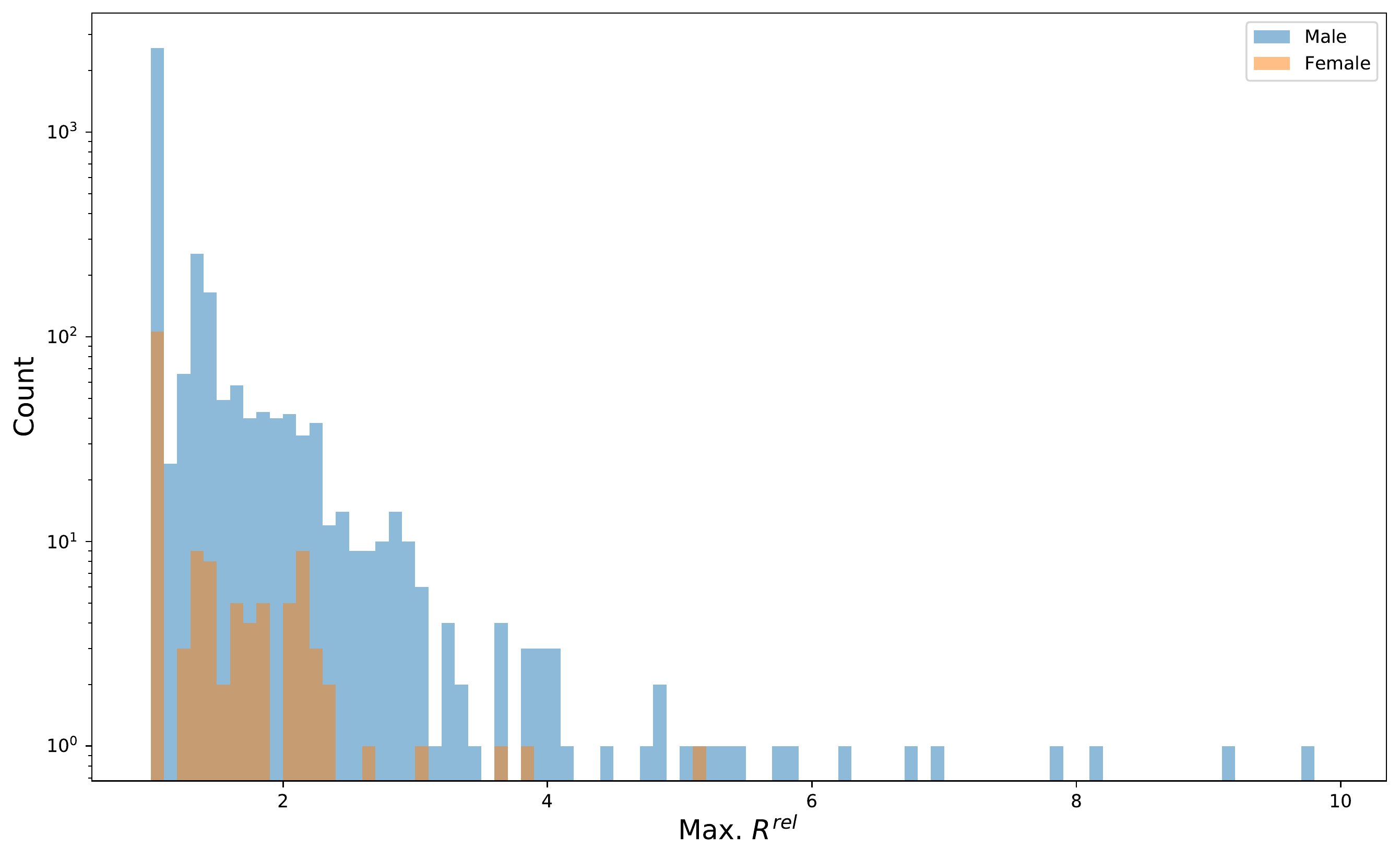}
      \caption{
        \emph{Histogram of maximum relative social amplification across names, separated by gender.} The bin size is 0.1. Although there are many more men than women in our analysis, their distributions of maximum relative social amplification are similar. The distributions skew right. There are some outliers (not shown), all men, with a maximum relative social amplification above 10.
      }
      \label{fig:maxratiohist}
    \end{figure*}

\end{document}